\documentclass[a4paper,11pt]{article}
\pdfoutput=1 
\usepackage{jcappub} 
\usepackage[T1]{fontenc} 
\usepackage[caption=false]{subfig}
\usepackage{graphicx}
\usepackage{epstopdf}
\usepackage{mathrsfs}
\usepackage{amssymb}
\usepackage{verbatim}
\usepackage{color}
\usepackage{multirow}
\usepackage{amsmath}
\usepackage{subfig}
\usepackage{lipsum}
\usepackage{hyperref}
\usepackage[utf8]{inputenc} 
\newcommand{\beq}{\begin{equation}}
\newcommand{\eeq}{\end{equation}}
\def\be{\begin{equation}}
\def\ee{\end{equation}}
\def\bea{\begin{eqnarray}}
\def\eea{\end{eqnarray}}
\def\mrm{\mathrm}
\def\lsim{\raise0.3ex\hbox{$\;<$\kern-0.75em\raise-1.1ex\hbox{$\sim\;$}}}
\def\gsim{\raise0.3ex\hbox{$\;>$\kern-0.75em\raise-1.1ex\hbox{$\sim\;$}}}

\subheader{LPT--Orsay 16-50, CPHT-RR044.082016}

\title{\boldmath Scrutinizing a di-photon resonance at the LHC  through Moscow zero}
\author[a]{Giorgio Arcadi}
\author[a,b]{Pradipta Ghosh,}
\author[a]{Yann Mambrini}
\author[a]{and Mathias Pierre}

\affiliation[a]{Laboratoire de Physique Th\'eorique, CNRS, Univ. Paris-Sud,\\
Universit\'e Paris-Saclay, 91405 Orsay, France}
\affiliation[b]{Centre de Physique Th\'eorique, \'Ecole Polytechnique, CNRS, \\
Universit\'e Paris-Saclay, 91128 Palaiseau, France}

\emailAdd{giorgio.arcadi@th.u-psud.fr}
\emailAdd{pradipta.ghosh@th.u-psud.fr}
\emailAdd{yann.mambrini@th.u-psud.fr}
\emailAdd{mathias.pierre@th.u-psud.fr}

\abstract{The ATLAS and CMS collaborations have recently released their new analyses
of the diphoton searches. We look in detail the consequences of their results deriving strong 
constraints on models where a scalar resonance $s$ decays into two light pseudoscalars which in 
turn decay into two pairs of collimated photons, mis-identified with two real photons. In our 
construction, all mass terms are generated dynamically, and only one pair of vector-like 
fermions generate couplings which will be probed using the upcoming LHC data. Moreover, we show that 
a stable dark matter candidate, respecting the cosmological constraints, is naturally affordable in the model.}


\arxivnumber{1608.04755}

\begin{document}
\maketitle
\flushbottom

\setcounter{equation}{0}


\section{Introduction}
\label{intro}


The detection of events with multi-photon final state at the LHC would represent one of the most 
distinct and accurate evidences of physics Beyond the Standard Model (BSM). The study of these kinds of 
events has encountered, in recent times, an increased interest from the Particle Physics community as a 
consequence of the announcement of a 3.9$\sigma$ (3.4$\sigma$) local excess in the diphoton channel 
by the ATLAS \cite{ATLASdiph,Moriondatlas} (CMS \cite{CMS:2015dxe,CMS:2016owr}) group, corresponding to 
a 750 GeV resonance with a cross-section $\sigma_{\gamma\gamma} \sim 4-16$ fb \cite{Moriondatlas}. 
One straightforward interpretation of this excess consists of the introduction of a spin-0 resonance 
coupled, at least to the gluons and the photons (as well as the $Z$ boson from gauge invariance) through 
dimension-5 operators of the form $\frac{c_{VV}}{\Lambda} V_{\mu \nu}V^{\mu \nu} \Phi$\cite{McDermott:2015sck,
Ellis:2015oso,Low:2015qep,Dutta:2015wqh,Falkowski:2015swt} with 
$\Lambda$ being the effective scale of these new interactions. The most widely studied origin of these 
dimension-5 operators is represented by new fermionic degrees of freedom, charged under color and hypercharge 
and  vector-like with respect to the Standard Model (SM)
~\cite{Angelescu:2015uiz}. These new degrees of freedom induce, 
through triangle loops, couplings between the 750 GeV resonance and gluon/photons which can be described through 
the aforementioned dimension-5 operators upon integrating out the new fermions.

Although the most recent measurements~\cite{ATLAS1,CMS1} seem not to confirm the existence of such a 
resonance (at least for a (pseudo-)scalar\footnote{Public announcement regarding a spin-2 resonance with latest data-set is
yet missing from ATLAS.} particle), 
scenario where spin-0 states are coupled with new fermions, having non-trivial 
quantum numbers under the SM gauge group, and detectable in diphoton events remains a natural and intriguing 
option to study. Indeed, as proven for example in refs.~\cite{Franceschini:2015kwy,
Gu:2015lxj,Salvio:2015jgu,Dev:2015vjd,Son:2015vfl,Bertuzzo:2016fmv,Salvio:2016hnf,Bae:2016xni,Hamada:2016vwk}, in order to achieve a 
production cross-section of the order of the current experimental sensitivity, possibly accompanied by a sizable decay width, 
one should require either (1) large and possibly non-perturbative, i.e., $> \sqrt{4\pi}$,
Yukawa couplings between the new BSM fermions and the 750 GeV resonance or (2) large 
number of new fermions or (3) large hypercharge assignment for these BSM fermions, unless rather particular 
setups are realized~\cite{Bharucha:2016jyr,DiChiara:2016dez,Djouadi:2016oey}. 
However, when one or more of these conditions are met then the behavior of the theory at scales 
above but not too far from the mass of the 
new fermions might become problematic. For example, condition (1) would cause a departure of the new Yukawa couplings, at 
low scales, from the perturbative regime while the last conditions would be associated to the 
occurrence of a Landau pole {\it{(a.k.a. Moscow zero)}}~\cite{lpole} for
the SM gauge couplings. Furthermore, the quartic coupling present in the new scalar sector, 
producing the diphoton 
resonance, would rapidly be driven to negative value, making the theory pathological.

It would then be intriguing to compare the current experimental limits 
with the requirement of having a consistent 
perturbative theory at least up to some energy scale $\Lambda_{\rm NP}$, where a further 
Ultra-Violet (UV) completion should be invoked. 
This problem can also be reformulated by choosing a priori the scale $\Lambda_{\rm NP}$ and determining the accessible 
values of an hypothetical future new detection, as function of the mass of the resonance 
and the diphoton production cross-section.

\begin{figure}
\begin{center}
 \includegraphics[width=0.5\linewidth]{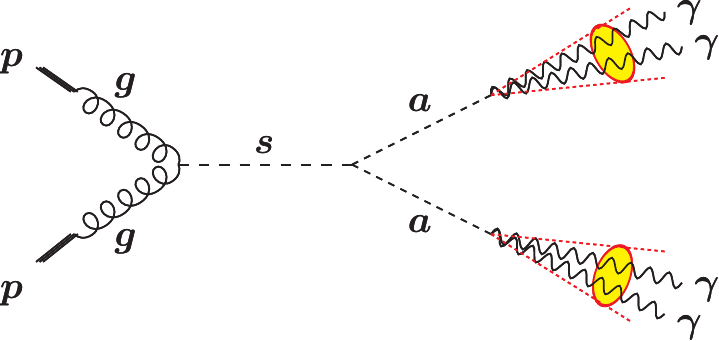}
 \caption{{\footnotesize Feynman diagram producing a 4-photon final state that is mis-identified 
 as a diphoton signal. 
 }}
\label{Fig:feynman}
\end{center}
\end{figure}

To perform a study of this kind we consider the case in which the SM is extended by a SM gauge singlet complex 
scalar field $\Phi$ charged under a new global $U(1)$ symmetry, which can be interpreted as a 
Peccei - Quinn symmetry for instance \cite{Aparicio:2016iwr}. This global symmetry is spontaneously broken by the vacuum 
expectation value (vev) of the scalar component of $\Phi$. The pseudoscalar 
component of $\Phi$ is promoted to be a light pseudo Goldstone boson by 
introducing small explicit violation of this new global symmetry. In this setup a diphoton signal would be 
originated from  pairs of highly collimated photons\footnote{A similar possibility
can appear in Hidden-Valley-like model~\cite{Chang:2015sdy}.}, produced by the 
decay of a pair of light pseudoscalars which in turn are originated 
by the decay of the scalar component of $\Phi$, resonantly produced in proton-proton collision 
(see figure~\ref{Fig:feynman}). 
The advantage of this construction is that the four-photon,
i.e., the effective diphoton, cross-section can more easily attain larger values 
since it is controlled by branching fraction (Br) of the 
decay of the scalar component of $\Phi$ into two pseudoscalars, 
which can easily be pushed to one. This would imply, in 
construction involving vector-like fermions for the generation of couplings with gluons and photons, a lower multiplicity 
for the additional fermions. Another interesting feature is that the complementary 
signals like dijets, $ZZ$ and $Z\gamma$ are generally suppressed, rendering the model easy to discriminate. 
A similar framework giving collimated photons is the Next-to Minimal Supersymmetric Standard Model
(NMSSM)~\cite{Ellwanger:2016qax, Domingo:2016unq,Badziak:2016cfd} in which the new scalar and 
pseudoscalar\footnote{Collimated photons from a light pseudoscalar in the NMSSM were also analysed 
earlier~\cite{Dobrescu:2000jt,Dobrescu:2000yn}.} fields are part of an 
extended Higgs sector. In this scenario a sizable production cross-section is achieved through triangle 
loop of the SM fermions, at the price of a very definite range of the pseudoscalar mass 
to avoid an otherwise highly suppressed branching fraction into photons.

The setup discussed above has previously been sketched in ref.~\cite{Arcadi:2016dbl} along with 
the introduction of a Dark Matter (DM) 
candidate. Interestingly, the mass term for the DM is dynamically generated by the vev of the field $\Phi$. 
This setup appears successful in fitting for instance the (now--dead) 750 GeV excess compatibly 
with the viable DM relic density without 
conflicting with bounds from DM detection, contrary to what would happened in the case of a ``true'' diphoton 
signal~\cite{Mambrini:2015wyu,Backovic:2015fnp,D'Eramo:2016mgv}.
In this work we will present a more complete anomaly--free and 
UV version of the model presented in ref.~\cite{Arcadi:2016dbl}
so that, similar to the DM interactions, the couplings among the field $\Phi$ and the SM gauge bosons 
are also dynamically 
generated. Indeed, the particle spectrum of the theory will be extended with new BSM fermionic states, vector-like with 
respect to the SM but chiral with respect to the new $U(1)$ symmetry, whose mass terms are generated by spontaneous 
breaking of the associated $U(1)$ symmetry. This kind of scenario is very predictive since 
it features, as free parameters, only masses of the new 
states (resonant scalar, new vector-like fermions
and the DM) and one fundamental coupling, i.e. the quartic coupling $\lambda$ of the $\Phi$ field.

We will show that asking for the stability of the $\Phi$ potential 
($\lambda$ positive up to a energy scale $\Lambda_{\rm NP}$, beyond with 
``further'' New Physics appears essential) 
provides constraints competitive with the experimental ones.
In particular, the masses of the new fermions are restricted to lie below 2 TeV, 
well within the kinematical 
reach of LHC run-II. By further combining the requirements of theoretical consistency and a
near future detection of the diphoton signal with correct relic density, 
an even more constrained scenario, giving definite predictions of the DM mass, is achieved.

The paper is structured as follows. We first discuss an overview of the model
in section \ref{Sec:model} and then provide the 
relevant expressions for the 4-photon production cross-section. 
In the following section we investigate the UV behavior 
of the theory by studying the relevant Renormalization Group Equations (RGEs). 
In section \ref{Sec:DM} we address the DM phenomenology. 
We summarize our results and discuss the associated implications in section
\ref{Sec:summary} before we put our concluding remarks in section \ref{Sec:conclu}.
Some useful formulas are relegated to the appendix.

\section{The Model}
\label{Sec:model}
\subsection{The Lagrangian}
\label{subsec:Lagrangian}

We minimally extend the Standard Model by introducing a complex scalar field $\Phi=(s +ia)/\sqrt{2}$, 
singlet under the SM gauge group, whose Lagrangian can be written as:
%
\beq
{\cal L}_\Phi = \partial_\mu \Phi \partial^\mu \Phi^* + \mu_\Phi^2 |\Phi|^2 - \lambda |\Phi|^4 
+ \frac{\epsilon_\Phi^2}{2} (\Phi^2 + \text{~h.c.}).
\label{Eq:lagrangian1}
\eeq

We can see from eq.~(\ref{Eq:lagrangian1}) that the last term ($\epsilon_\Phi \ll \mu_\Phi$) 
explicitly breaks the $U(1)$ symmetry, giving mass to the otherwise massless Goldstone boson $a$, 
which is then promoted to a light pseudo-Goldstone state\footnote{As it was shown 
in refs.~\cite{dudas1,dudas2}, non--perturbative effects can give mass to 
the Goldstone boson, generating a large hierarchy between 
$\mu_\Phi$ and $\epsilon_\Phi$ through instantonic effects.}. We also 
introduce new Dirac fermions $F_L,F_R$, chiral under the new $U(1)$ symmetry with different charge assignments
but vector-like under the SM gauge group, having non trivial  $SU(3)_C$ and $U(1)_Y$ quantum numbers
but singlet with respect to $SU(2)_L$. 
With these assumptions the relevant Lagrangian for the new fermions is:

\beq
\label{Eq:lagrangian3}
{\cal L}_F= i  \bar{F}_L \gamma^\mu D_\mu F_L+ i  \bar{F}_R \gamma^\mu D_\mu F_R - (y_F \Phi \bar{F}_L F_R 
+ \text{h.c.}) = i \bar{F} \gamma^\mu D_\mu F - \frac{y_F}{\sqrt{2}} s \bar{F} F
- i \frac{y_F}{\sqrt{2}} a \bar{F} \gamma^5 F,
\eeq
where $D_\mu$ is the SM covariant derivative and $F \equiv F_L+F_R$.

Once $\Phi$ acquires a vev through the spontaneous 
symmetry breaking mechanism, $\Phi= (v_\Phi + s + i a)/\sqrt{2}$ with 
$v^2_\Phi = (\mu_\Phi^2 + \epsilon^2_\Phi)/\lambda
\simeq \mu_\Phi^2 /\lambda$,
it generates automatically the mass of $s$ ($m_s=\sqrt{2 \lambda} v_\Phi$), the
$saa$ coupling ($\lambda_{saa}= \lambda v_\Phi$), and masses of the new 
fermions ($m_F= y_F v_\Phi/ \sqrt{2}$). The pseudoscalar
mass is given by $m_a=\sqrt{2}\epsilon_\Phi$.
Re-expressing the Lagrangian, after spontaneous symmetry breaking, as a function 
of the physical fields gives :
%
 \beq
\label{Eq:lagrangian4}
 -{\cal L} \supset  \frac{m_s^2}{2} s^2 + \frac{m_a^2}{2}a^2 + \sqrt{\frac{\lambda}{2}} m_s sa^2 
 + \sqrt{\frac{\lambda}{2}} m_s s^3  + \frac{\lambda}{4} (s^2+a^2)^2 
+ m_F \bar F F + \sqrt{2 \lambda} \frac{m_F}{m_s} 
 (s \bar F F + i a \bar F \gamma^5 F),
 \eeq
where we have used $y_F=\sqrt{2}m_F/v_\Phi=2\sqrt{\lambda}(m_F/m_s)$. 
We note in passing that in the current article, for simplicity and enhanced predictability, 
we have assumed vanishing mixing between $\Phi$ and the SM-Higgs. This choice is also motivated 
by the fact that this mixing is in general strongly constrained. For example, it would lead to 
non-standard decays of the SM-Higgs, e.g., SM-Higgs to a pair of light pseudoscalars. Similarly, 
a sizable mixing between $\Phi$ and the SM-Higgs would appear strongly disfavored by the DM phenomenology. 
Indeed it would induce a coupling between the DM and the SM-Higgs, responsible for a potentially 
strong Spin Independent component of the DM scattering cross-section with nucleons, which 
would be in tension with the current experimental limits~\cite{Akerib:2015rjg}.
 
\subsection{Generating the vertices}
\label{subsec:vertices}

\begin{figure}
\begin{center}
 \includegraphics[width=0.6\linewidth]{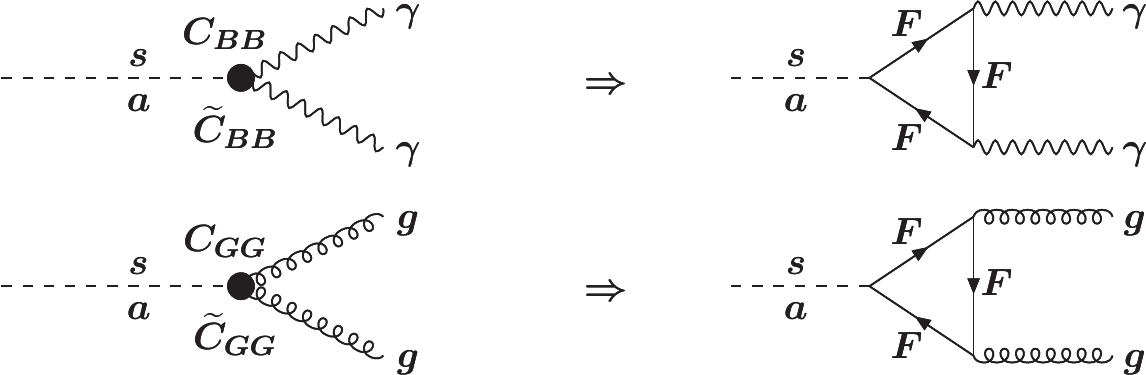}
 \caption{{\footnotesize Loop triangle diagram generating 
 interactions among the photons and gluons with the 
 real $(s)$ and imaginary $(a)$ components of $\Phi$. 
 The first diagram, from the principle
 of gauge invariance, will also generate
 effective interactions like $sZ\gamma,\,sZZ$, etc, however,
 their relative strengths will be different from $s\gamma\gamma$ \cite{Arcadi:2016dbl}.
 }}
\label{Fig:triangle}
\end{center}
\end{figure}

Effective couplings between the SM gauge fields and the complex scalar field $\Phi$ are generated through a 
triangular loop (see figure~\ref{Fig:triangle}) involving the new fermions, $F$. In this work the new fermion 
sector will be composed by $N_f$ ``families''; each one
comprised of a pair of degenerate fermions, chiral under the new $U(1)$ symmetry
and belonging to the fundamental representation of $SU(3)_C$, singlet under $SU(2)_L$ 
and, with electric charges $\pm Q_{F}$ (i.e., the $U(1)_Y$ charges), in order to represent 
an anomaly free configuration. 
We observe that the main purpose of this paper is to provide a 
proof of existence of the correlation between hypothetical signals of collimated photons, 
UV consistency of the underlying theory and DM phenomenology. As a consequence, we have presumed some 
simplifying assumptions. We have indeed considered, throughout this work, that both the 
production process (gluon fusion) and the decay of the scalar resonance into $4 \gamma$ are originated by the 
same fermion fields, which are, thus, both electrically and color charged. As further assumption we have 
considered only integer values for $Q_F$. Our results could be straightforwardly extended 
to more generic assumptions concerning the parameters of the new fermionic sector. 
It is worthy to note here that strong experimental bounds exist on the masses of vector-like 
quarks \cite{Aad:2015kqa,Khachatryan:2015oba}, on exotic particles having higher electric charges $(2|e|-6|e|)$
\cite{Aad:2015oga} or on lepton-like fermions with $Q_F=1|e|,\,2|e|$ \cite{Khachatryan:2016sfv}.
However, experimental constraints on colored vector-like fermions having integral electric charges,
as considered in this article, remain rather inadequate till date.

\noindent
In this setup the relevant Lagrangian can be expressed as \cite{Arcadi:2016dbl}: 
\beq
\label{Eq:lagrangian5}
 -{\cal L}  \supset \frac{\sqrt{\lambda} C_{GG}}{m_s} s G_{\mu \nu}^\alpha G^{\mu \nu}_\alpha
 + \frac{\sqrt{\lambda} \widetilde C_{GG}}{m_s} a G_{\mu \nu}^\alpha \widetilde G^{\mu \nu}_\alpha 
+ \frac{\sqrt{\lambda} C_{BB}c_W^2}{m_s} s F_{\mu \nu} F^{\mu \nu}
 + \frac{\sqrt{\lambda} \widetilde C_{BB}c_W^2}{m_s} a F_{\mu \nu} \widetilde F^{\mu \nu},
\eeq
where $F_{\mu \nu}$ and $G_{\mu \nu}^\alpha$
($\widetilde F_{\mu \nu}$ and $\widetilde G_{\mu \nu}^\alpha$) are the 
field (dual field) strengths for the photon and gluons, respectively.
Here $c^2_W\equiv \cos^2\theta_W=0.769$ \cite{Agashe:2014kda}, with $\theta_W$ being the Weinberg angle,
and the coefficients (details are given in the appendix):
\bea
\label{eq:coeff}
&& C_{GG} = N_f\frac{\alpha_s}{4 \sqrt{2} \pi} f_{1/2}(m_s^2/4 m_F^2),
\,\,\, C_{BB} =N_f \frac{3 \alpha_{em}}{2 \sqrt{2} \pi c_W^2} Q^2_F f_{1/2}(m_s^2/4 m_F^2), \nonumber\\
&&\widetilde C_{GG} =N_f \frac{\alpha_s}{4 \sqrt{2} \pi} \tilde f_{1/2}(m_a^2/4 m_F^2),
\,\,\, \widetilde C_{BB} = N_f\frac{3 \alpha_{em}}{2 \sqrt{2} \pi c_W^2} Q_F^2 \tilde f_{1/2}(m_a^2/4 m_F^2),
\eea

\noindent
where $\alpha_s,\,\alpha_{em}$ are the strong and electromagnetic coupling
constants and the functions $f_{1/2}(m_s^2/4 m_F^2)$, $\tilde{f}_{1/2}(m_a^2/4 m_F^2)$
are described in the appendix. Note that the presence of $Q^2_F$
in the expressions for $C_{BB},\,\widetilde C_{BB}$ makes them insensitive to the sign of $Q_F$.
It is evident from eq.~(\ref{Eq:lagrangian5}) that
the scale of dimension-5 operators $\Phi V^{\mu \nu} V_{\mu \nu}$ 
corresponds to the vev $v_\Phi \propto {m_s}/{\sqrt{\lambda}}$ of the complex scalar field. The coefficients
in eq.~(\ref{eq:coeff}) 
are substantially insensitive\footnote{Apart from a mild dependence from $f_{1/2}(m_s^2/4 m_F^2)$,
$\tilde{f}_{1/2}(m_a^2/4 m_F^2)$ functions for $m_s^2/4 m_F^2,m_a^2/4 m_F^2< 1$ \cite{Angelescu:2015uiz}.}
 to $m_s$, $m_a$ and the masses of new fermions $m_F$.
From now on -unless differently stated- we will consider, throughout this work, 
the case of only one pair of new fermions, i.e., $N_f=1$.

\subsection{The 4$\gamma$ production}
\label{subsec:4gamma}
In this scenario, the diphoton signal is mostly originated from the process 
$gg \rightarrow s \rightarrow aa \rightarrow 4 \gamma$, whose corresponding diagram is shown in 
figure~\ref{Fig:feynman}. Given a sufficiently light pseudoscalar, the boosted photon pairs originated from 
its decay might appear enough collimated to be mis-identified as a single photon.

\noindent
As minimal requirement to achieve such degree of collimation, one needs to impose that the opening angle 
between the collimated photons remains below the detector resolution of the LHC, $\sim\mathcal{O}($20 mrad). 
This can be accomplished by imposing the boost factor $\gamma \simeq \frac{m_s}{2 m_a} 
\gtrsim 200$~\cite{Chala:2015cev} implying:
\beq
m_a \lesssim 2.5~\mrm{GeV} \left( \frac{m_s}{1~\mrm{TeV}} \right),
\label{Eq:mamax}
\eeq
and thus, obtaining an upper bound on the mass of the pseudoscalar for a fixed $m_s$. 
On the other hand, such a low mass might produce 
large lifetime for the pseudoscalar field such that it decays beyond the Electromagnetic CALorimeter 
(ECAL) which lies at a distance of $\mathcal{O}(1$ m) from the primary interaction point. 
Such situation mainly arises for $m_a \lesssim 3 m_{\pi_0}$ since in this 
case the pseudoscalar can only decay into photon pairs. In this setup, 
requiring a decay length $l \lesssim 1\,\mbox{m}$ translates into the following relation:

\beq
l \simeq 71 ~ \mrm{m} \left( \frac{m_s}{1~\mrm{TeV}} \right)^3 \left( \frac{0.5 ~\mrm{GeV}}{m_a} \right)^4
\frac{1}{N^2_f Q_F^4} \left(  \frac{0.01}{\lambda}\right),
\label{Eq:mamin}
\eeq
which can be used to extract a lower bound on $m_a$. In the case of $m_s=750\,\mbox{GeV}$ for instance, 
the combination 
of eq.~(\ref{Eq:mamax}) and eq.~(\ref{Eq:mamin}) fixes the value of $m_a$ to lie within the range 
$0.2$ GeV $\lesssim m_a \lesssim$ 2 GeV \cite{Arcadi:2016dbl,Knapen:2015dap,Aparicio:2016iwr}.
This mass range could be restricted further by a more dedicated 
investigation of the ECAL response~\cite{Knapen:2015dap,Aparicio:2016iwr} and/or by including the effects of photon 
conversions~\cite{Dasgupta:2016wxw}. These kinds of analyses are, however, beyond the scopes of this paper.

The 4$\gamma$ production cross-section, following figure~\ref{Fig:feynman}, can be straightforwardly computed as: 

\beq
\label{4gammma1}
\sigma_{4 \gamma} = \frac{\pi^2}{8 m_s \bf{s}} \Gamma(s \rightarrow gg) \mrm{Br}(s \rightarrow aa) 
[\mrm{Br}(a \rightarrow \gamma \gamma)]^2 I_{GG}(m_s/\sqrt{\bf{s}}),
\eeq
where $\sqrt{{\bf s}}=13\,\mbox{TeV}$ is the present LHC center-of-mass energy,
$\Gamma(s \rightarrow gg)$ is the decay width of $s\to g g$ process
and $I_{GG}(m_s/\sqrt{\bf{s}})$ is the integral representing 
the effect of the parton distribution functions (PDFs). For the computation 
of $I_{GG}(m_s/\sqrt{\bf{s}})$ we have used the 
MSTW2008NNLO distribution~\cite{Martin:2009iq,Martin:2009bu,Martin:2010db}.

It can be easily verified that, in the parameter space relevant for $gg\to s\to aa\to 4\gamma$ 
process, $\mrm{Br}(s \rightarrow aa) \sim 1$ \cite{Arcadi:2016dbl}. As already 
pointed out, for $m_a < 3 m_{\pi^0}$, only decay into the photon pairs is accessible, so the 4$\gamma$ 
production cross-section depends 
only on the coupling $C_{GG}$ (contained in $\Gamma(s \rightarrow gg)$) between the resonance and the gluons.
Thus, eq.~(\ref{4gammma1}), using eq.(\ref{eq:coeff}), can be simplified as:

\bea
\sigma_{4 \gamma} \simeq 1.64~\mrm{fb} \frac{(\Gamma_s/m_s)}{10^{-4}} 
\left( \frac{I_{GG}(m_s/\sqrt{\bf{s}})}{2000} \right) N^2_f
\,
 \simeq 0.33~\mrm{pb}\left( \frac{ I_{GG}(m_s/\sqrt{\bf{s}})}{2000} \right) N^2_f \lambda,\nonumber\\
\hspace*{9cm}\left[\rm{for~} m_a \lesssim 3 m_{\pi^0}\right],~~
\label{Eq:lmpi}
\eea	
here $\Gamma_s$ represents the total width of scalar resonance $s$,
which is $\approx \Gamma(s\to aa)$ when diphoton signal
is arising from collimated photons. For numerical estimates
we have used values of $\alpha_{em},\,\alpha_s$ at the energy
scale $m_Z$ \cite{Agashe:2014kda}.

In the case of $m_a \gtrsim 3\,m_{\pi^0}$, the production cross-section should be rescaled by a factor $\mrm{Br}(a 
\rightarrow \gamma \gamma)^2 \approx \frac{81}{4} \frac{\alpha_{\rm em}^4}{\alpha_{\rm s}^4}Q_F^8$ 
and\footnote{This expression for $\mrm{Br}(a \rightarrow \gamma \gamma)^2$ holds true for 
$Q_F\leq 4$.
For larger $Q_F$, Br$(a\to \gamma\gamma)$ is again $\approx 1$. However,
when $\sqrt{N_f}Q_F > 4$, Br$(s\to aa)$ is no longer $\approx 1$.} is given by:
\bea
\label{Eq:gmpi}
\sigma_{4\gamma}\simeq 0.63~\mrm{fb} \frac{(\Gamma_s/m_s)}{0.1} 
\left( \frac{ I_{GG}(m_s/\sqrt{\bf{s}})}{2000} \right) N^2_f Q^8_F \simeq 0.12~\mrm{fb} \,\,
\left( \frac{ I_{GG}(m_s/\sqrt{\bf{s}})}{2000} \right) N^2_f Q^8_F\,\lambda, \nonumber\\
\hspace*{8cm} \left[\rm{for}~m_a \gtrsim 3 m_{\pi^0}\right].\,\,
\eea

Just like the pseudoscalar field, the scalar field $s$ is also
coupled to photons through the same set of new BSM fermions
(see figure~\ref{Fig:triangle}). Thus, one should in general also
account for the direct diphoton production from the decay of scalar resonance. 
The conditions for the dominance of  
4-photon production cross-section over the 
2-photon $(\sigma_{2\gamma})$ have already been determined in ref.~\cite{Arcadi:2016dbl} 
in terms of the effective coupling coefficients. In the dynamical realization, considered in 
this work, these conditions can be translated as\footnote{In these derivations
we have assumed that $f_{1/2}(m^2_s/4m^2_F)$ is behaving as $\tilde f_{1/2}(m^2_a/4m^2_F)$
for $m^2_s/4m^2_F$, $m^2_a/4m^2_F <1$.}:
\bea
&& C_{\rm BB} < 0.65 \rightarrow \sqrt{N_f} Q_F \lesssim 11,\,\,\,\,\,\,\,\left[\rm{for}~m_a < 3 m_{\pi^0}\right], \nonumber\\
&& \widetilde C_{\rm GG} < 0.27 \widetilde C_{\rm BB}/\sqrt{ C_{\rm BB}}
\rightarrow  Q_F/\sqrt{N_f} \lesssim 1 \,\,\,\,\,\,\,\left[\rm{for}~m_a > 3 m_{\pi^0}\right],
\eea
while the ``true'' diphoton production cross-section $\sigma_{2\gamma}$
can be written as: 
\beq
\label{2gammma1}
\sigma_{2 \gamma} = \frac{\pi^2}{8 m_s \bf{s}} \Gamma(s \rightarrow gg) \mrm{Br}(s \rightarrow \gamma\gamma) 
I_{GG}(m_s/\sqrt{\bf{s}}) \,\, \simeq 3.7\times 10^{-4}~\mrm{fb}\,\, \frac{N^4_f Q^4_F}{(\Gamma_s/m_s)} 
\left( \frac{I_{GG}(m_s/\sqrt{\bf{s}})}{2000} \right) \lambda^2.
\eeq

We notice that in the regime $m_a < 3 m_{\pi^0}$, the contribution to the collective 
diphoton, i.e., $2\gamma+4\gamma$, production from 
$s\to \gamma\gamma$ process is practically 
negligible unless one considers very high fermion multiplicities and/or high electric charges. In the opposite regime, 
in the case of $N_f=1$ and $Q_F \leq 2$, which will be considered throughout most of this work, a sub-dominant but 
non-vanishing contribution from the $gg \rightarrow s \rightarrow \gamma \gamma$ process is present. 
Although the analytical estimates presented in the current and the following subsection 
will rely only on the 4-photon production cross-section, nevertheless,
in our numerical analysis we have also included the 2-photon production cross-section.

One can easily see from eq.~(\ref{Eq:lmpi}) and eq.~(\ref{Eq:gmpi}), 
that the 4-photon production cross-section is mostly determined by the parameters $\lambda$ 
(and $Q_F$ for $m_a > 3 m_{\pi^0}$) while the dependence on $m_s$ is encoded only in the 
integral $I_{\rm GG}(m_s/\sqrt{\bf{s}})$. It is then possible to translate the 
observed limits by the ATLAS  \cite{ATLAS1} and CMS \cite{CMS1}
collaborations on the diphoton production cross-section directly
into limit on the value of $\lambda$. These would particularly give
stringent bound on $\lambda$ in the $m_a \lesssim 3 m_{\pi^0}$ case, for example, 
$\lambda \lesssim 3 \times 10^{-2}$ for $m_s=1$ TeV, 
using the latest experimental result \cite{CMS1}.

By comparing eq.~(\ref{Eq:lmpi}) in the light of eq.~(\ref{Eq:mamin}), we also notice that the requirement of a
decay length for the pseudoscalar $\lsim 1$ m would imply a lower limit on the 4$\gamma$ as: 

\beq
\sigma_{4\gamma} \gtrsim 232\, \mrm{fb} \frac{1}{Q_F^4} 
\frac{I_{GG}(m_s/\sqrt{\bf{s}})}{2000} \left( \frac{m_s}{1~\mrm{TeV}} \right)^3 
\left( \frac{0.5~\mrm{GeV}}{m_a} \right)^4,
\eeq
which, given the strong constraint on the observed diphoton cross-section,
will severely restrict or even discard the $m_a \lesssim 3 m_{\pi^0}$ scenario.
\begin{figure}[t!]
\begin{center}
\includegraphics[width=7 cm]{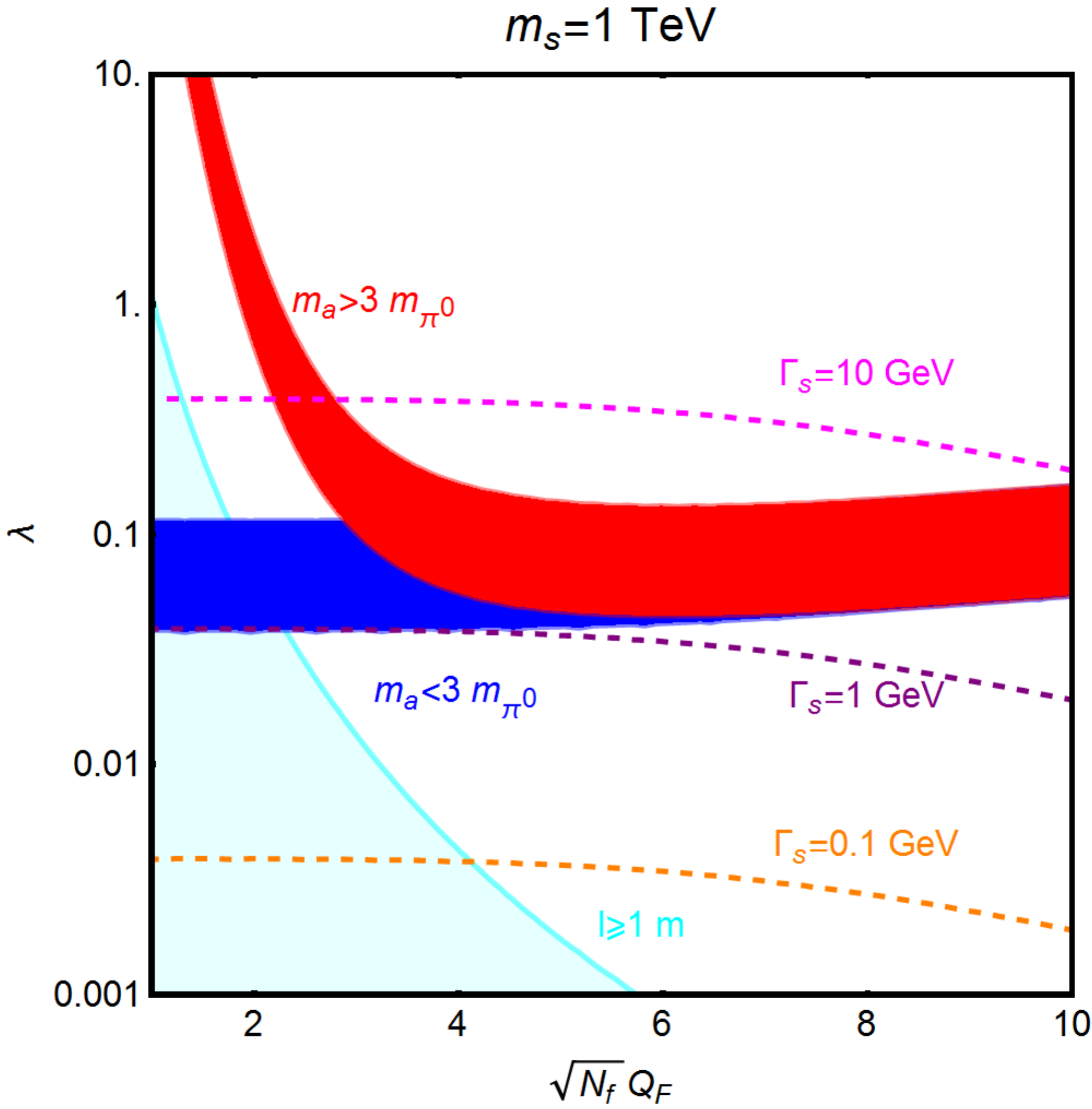}
\end{center}
\caption{\footnotesize{Colored contours representing the different values
of $\Gamma_s$ in the bi-dimensional plane $(\sqrt{N_f}Q_F, \lambda)$ corresponding to 
$2$ fb $\lesssim \sigma_{4\gamma}\lesssim 6$ fb for $m_s=1\,\mbox{TeV}$
with $m_a < 3 m_{\pi^0}$ (blue colored contour) and $m_a > 3 m_{\pi^0}$ (red colored contour). The 
cyan colored region in the left, where the pseudoscalar 
decays beyond 1 m, is excluded. }}
\label{fig:magL}
\end{figure}

A sketch of the aforesaid scenario is depicted in figure~\ref{fig:magL}. 
Keeping $m_s$ fixed at 1 TeV, we can see from figure~\ref{fig:magL} that 
in the $m_a < 3 m_{\pi^0}$ case, a sizable production cross-section can be obtained, compatible with $l \lsim 
1\,\mbox{m}$, only for $Q_F \gtrsim 4$ (or with a very high family number $N_f$). 
The variation of $l$
with $\sqrt{N_f}Q_F$ and $\lambda$ is consistent with eq.~(\ref{Eq:mamin})
which predicts smaller pseudoscalar decay lengths
for increasing $\sqrt{N_f}Q_F$ and $\lambda$ values. On the contrary, the limit on 
the decay length does not affect the $m_a > 3 m_{\pi^0}$ scenario since one has to consider much larger values of $\lambda$ 
to compensate (i.e., through Br$(s\to a a )$) the suppression from $\mrm{Br}(a \rightarrow \gamma \gamma)^2$
as long as $Q_F\leq 4$. We notice that for $Q_F \gtrsim 4$ 
the predictions for the two regimes, $m_a > 3 m_{\pi^0}$ and $m_a < 3 m_{\pi^0}$, tend to coincide
since for such large $Q_F$ values, the quantity Br$(a\to\gamma\gamma)$ remains
$\approx 1$ irrespective of the fact whether the hadronic channel is kinematically accessible or not.

It is interesting to note that for this model, due to its dynamical construction, a direct
link between the width of the resonance and the multi-photon production cross-sections
$(\sigma_{2\gamma},\,\sigma_{4\gamma})$ appears feasible. Hence, given this model,
any future measurement of an excess by the ATLAS or CMS group could be used
to estimate the total width as given by eq.~(\ref{Eq:lmpi}) and eq.~(\ref{Eq:gmpi}).
We note in passing that the results derived up to now are substantially insensitive to the mass of the new 
fermions. However, as will be discussed in the next section, the mass $m_F$ plays a very relevant role once the UV 
consistency of the theory is taken into account and this will translate, in turn, into stringent constraints 
on the low-energy phenomenology.

\section{The ultra-violet regime}
\label{Sec:UV}

As already stated in the introduction, the SM model extensions with new fermionic 
states, having sizable $(\mathcal{O}(1))$ couplings with the new
scalar (pseudoscalar) fields, face potentially strong constraints from the RGE evolution of 
the different relevant parameters of the theory. For example,
modified $\beta$ functions of the SM gauge couplings,
in the presence of new BSM states, can possibly lead 
to a Landau pole rather fast. As already argued in ref.~\cite{Goertz:2015nkp}, the parameter which 
is highly sensitive to radiative corrections is the quartic coupling $\lambda$, whose $\beta$ function contains 
a negative contribution proportional to $y_F^4$. Hence, a starting value of $y_F \sim \mathcal{O}(1)$
or higher, as typically required for having sizable diphoton cross-sections, will drive the quartic coupling 
towards negative values at a rather low energy scale. Thus, the vacuum of the theory
gets destabilized unless additional and opposite signed contributions are added
at the cost of incorporating more new bosonic degrees of freedom.

In our dynamical framework, the Yukawa couplings of the new fermions directly 
depend on their masses, the 
mass of the scalar resonance $m_s$ as well as the quartic coupling $\lambda$ :
$y_F = 2 \sqrt{\lambda} m_F/m_s$. Parameters $m_s$ and $\lambda$ being
responsible for the diphoton production cross-section, it appears feasible to connect 
the experimental limits on the  diphoton production cross-section and the requirement of the UV consistency 
of the theory, depending on the mass of the vector-like fermions.

Indeed, the relevant RGE equations for the studied dynamical model are given by:
\bea
&& \frac{d{y}_F}{d\ln\mu}=\beta_y =\frac{1}{16 \pi^2} \left((1+6 N_f )  {y}_F^3- 24 \pi {y}_F \alpha_1 
Q^2_F-32 \pi {y}_F \alpha_s  \right),\nonumber\\ 
&& \frac{d \lambda}{d \ln \mu} = \beta_\lambda = \frac{1}{16 \pi^2} 
(20 \lambda^2 - 12 N_f y_F^4 + 24 \lambda N_f y_F^2),
\label{Eq:betalambda}
\eea
where $\alpha_1 = g_1^2/4 \pi$ and $\alpha_s = g_3^2/4 \pi$. 
At a energy scale $\mu \geq m_F$, the gauge couplings $\alpha_1,\alpha_s$ are given by:
\bea
\label{eq:gauge}
&& \alpha_1(\mu)={\left[\frac{1}{\alpha_{1,\rm SM}(m_F)}-\frac{b_1^{\rm SM}+\Delta b_1}{2 \pi}
\ln\left(\frac{\mu}{m_F}\right)\right]}^{-1}, \nonumber\\
&& \alpha_s(\mu)={\left[\frac{1}{\alpha_{s,\rm SM}(m_F)}-\frac{b_3^{\rm SM}+\Delta b_3}{2 \pi}
\ln\left(\frac{\mu}{m_F}\right)\right]}^{-1},
\eea
where $(b_1^{\rm SM},b_3^{\rm SM}, \Delta b_1, \Delta b_3)=({41}/{6},-7,8 N_f Q_F^2, 
{4}N_f/3)$~\footnote{As pointed out in refs.~\cite{Bae:2016xni,Alves:2014cda}, 
the $\Delta b_i$ values considered 
in this work are consistent with the current LHC limits
and would appear detectable with higher luminosity or larger centre-of-mass energy.}. 
A Landau pole is considered when at a given energy scale $\mu,$
$\alpha_1(\mu),\,\alpha_s(\mu)=1$. Given the chosen gauge quantum number assignments of the new fermions, 
the coupling $\alpha_1$ is the one which might most 
likely encounter a Landau pole first. However, for $N_f=1$ and $Q_F \leq 2$, 
as considered for most of our analysis (see next section for clarification), the Landau pole 
of $\alpha_1$ lies at rather high scales, above $10^9\,\mbox{GeV}$. 
As a consequence, the constraints derived from Renormalization Group (RG) evolution 
apply mainly for the parameters $y_F$ and $\lambda$, as described by eq~.(\ref{Eq:betalambda}).
The quartic coupling $\lambda$, as also discussed in ref.~\cite{Goertz:2015nkp}, is very sensitive to the
radiative corrections. A simple analytical estimate can then be obtained as suggested in ref.~\cite{Goertz:2015nkp} 
by inspecting the ratio $|\beta_\lambda/\lambda|$. A value of this ratio greater than one would imply a very fast 
variation of $\lambda$ with energy scale, triggered by the radiative corrections, 
such that the theory would show a pathological behavior 
at rather low energy scales, unless suitably supplemented with new
contributions. In particular we notice that $\beta_\lambda$ contains a negative 
contribution scaling as $y_F^4$ with a rather large coefficient. Hence, for a high enough starting value of $y_F$ (in this 
case $\beta_y >0$ and so $y_F$ would increase with the energy scale\footnote{This would also imply that 
the parameter $y_F$
gets driven to the non-perturbative regime at some energy scale. In reality, however,
the parameter $\lambda$ attains negative values simultaneously
or slightly before, making the theory already ill-behaved.}) 
the quartic coupling $\lambda$ would be driven towards a negative value, 
destabilizing the vacuum of the theory. 
By approximating $\beta_\lambda \approx -\frac{12}{16 \pi^2} N_f y_F^4$ 
from eq.~(\ref{Eq:betalambda}) and substituting $y_F = 2 \sqrt{\lambda} 
\frac{m_F}{m_s}$ one obtains: 
\beq
\label{eq:estimate1}
\left|\frac{\beta_\lambda}{\lambda}\right| \simeq \frac{12 N_f}{\pi^2} \lambda \left( \frac{m_F}{m_s} \right)^4,
\eeq
which gives $|\beta_\lambda/\lambda| \simeq 1.2 \lambda \left( \frac{m_F}{m_s} \right)^4$ for $N_f=1$.

Now one can trade $\lambda$ with $\sigma_{4 \gamma}$ to translate eq.~(\ref{eq:estimate1})
into an upper bound for the diphoton cross-section. Considering eq.~(\ref{Eq:gmpi}) for example
with $N_f=1$, one obtains\footnote{Although eq.~(\ref{eq:estimate1}) and eq.~(\ref{eq:estimate2}) turn out 
to be good approximations, 
we have used them only for illustrative purpose. Our numerical results are based on the 
solution of the RGEs.}: 
\beq
\label{eq:estimate2}
\sigma_{4 \gamma} \lesssim 0.83\,{\mbox{fb}} \,\, Q_F^8 \left( \frac{m_s}{m_F} \right)^4 
\left( \frac{I_{GG}(m_s/\sqrt{s})}{2000} \right).
\eeq

For $N_f=1$ and $Q_F=2$, one obtains  $\sigma_{4 \gamma} \lesssim 0.5$ fb  
($0.07$ fb) for $m_F= 3$ TeV (5 TeV) in the case of a $m_s=1$ TeV resonance, 
which are well below the current experimental sensitivity \cite{ATLAS1,CMS1}. 
This behavior can be easily understood by the 
fact that larger $m_F$ (i.e., larger $y_F$) should be 
compensated with smaller $\lambda$ to maintain
$|\beta_\lambda/\lambda|$ well behaved, i.e., $<1$ (see eq.~(\ref{eq:estimate1})).
In other words, an observation of a signal 
in the near future would imply that the mass of the new fermions should not be too far from the one of the 
resonance and their future detection are well envisaged.

For numerical studies, we have performed a more systematic investigation by solving the 
relevant RGEs\footnote{Our 
analysis provide a proof of principle of the strong relation between the UV 
behavior of the theory and the low-energy phenomenology within a dynamical construction. However, our 
estimation of $\Lambda_{\rm NP}$ would differ when one considers higher order $\beta$ 
functions beyond the one-loop.}, i.e., eq.~(\ref{Eq:betalambda}) in combination with eq.~(\ref{eq:gauge}).   
%
\begin{figure}[t!]
\begin{center}
 \includegraphics[width=0.5\linewidth]{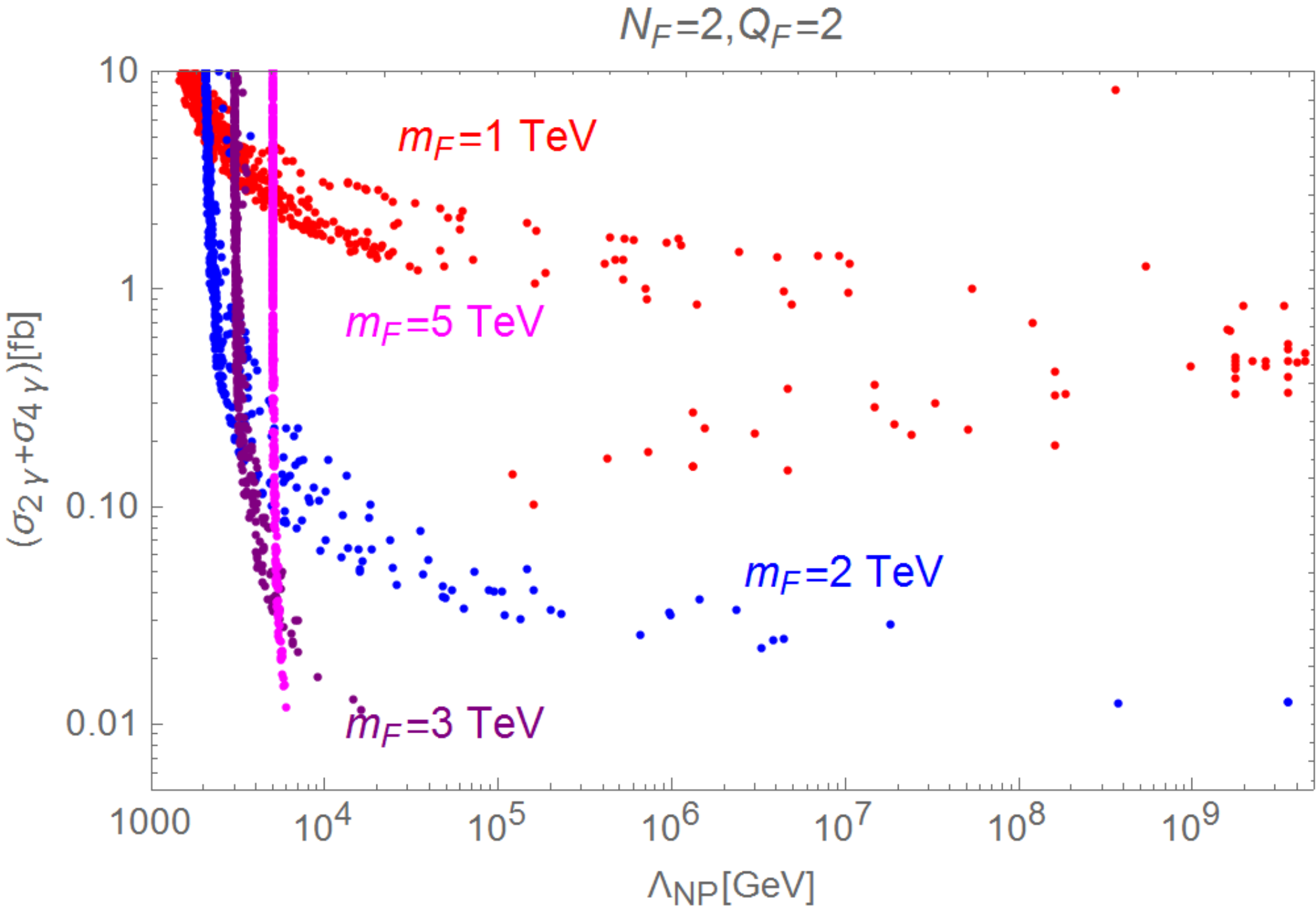}
 \caption{{\footnotesize The collective production cross-section 
 for a diphoton-like signal as a function of the scale $\Lambda_{NP}$ 
 where the quartic coupling becomes negative. Here, we consider $N_f=1,Q_F=2$ and the 
 different $m_F$ values are mentioned in the plot.}}
\label{Fig:lambdanp}
\end{center}
\end{figure}
%
The result of our numerical study is illustrated by figure \ref{Fig:lambdanp}. 
which represents the scan on
the parameters space $\lambda,m_s$ and $m_a$ (see section \ref{Sec:summary} for details) 
for different values for $m_F$, namely 
$1,2,3$ and $5$ TeV with $N_f=1,\,Q_F=2$. 
Figure \ref{Fig:lambdanp} depicted the collective value of diphoton
production cross-section, i.e., $\sigma_{2 \gamma} + \sigma_{4 \gamma}$
as a function of the scale $\Lambda_{NP}$ where the quartic coupling $\lambda$ becomes negative. We can 
clearly see in the figure that for 
 $m_F \leq 2$ TeV it is possible to have sizable diphoton production cross-section up to $\Lambda_{\rm NP} 
\sim O(10^9\,\mbox{GeV})$, a value also corresponding to the Landau pole of $\alpha_1$. On the contrary, all the  
configurations with $m_F=3-5\,\mbox{TeV}$ feature $\Lambda_{\rm NP} \sim m_F$. This implies that for these masses 
new degrees of freedom should already be introduced at scales equal or below $m_F$ to secure
a well-behaved theory up to some energy scale much larger than the involved mass spectrum. 

\section{Introducing a dark sector}
\label{Sec:DM}

It is rather straightforward to embed Dark Matter (DM) candidates in this dynamical framework by adding 
extra fermions charged only under the new $U(1)$ symmetry but singlet under the SM gauge group.
In this paper we will consider the case of a Majorana fermion $\chi$ described by the following Lagrangian: 

\beq\label{Eq:lagrangiandm}
{\cal L}_\chi =\frac{1}{2} i \bar \chi \gamma^\mu \partial_\mu \chi - y_\chi \Phi \bar \chi^c \chi 
+ \text{h.c.} \, \, \,\,
= \frac{1}{2} i \bar \chi \gamma^\mu \partial_\mu \chi - \frac{y_\chi}{\sqrt{2}} s \bar \chi^c \chi 
- i \frac{y_\chi}{\sqrt{2}} a \bar \chi^c \gamma^5 \chi.
\eeq

Similar to the new BSM fermions, producing interactions among
$\Phi$ with photons and gluons, the DM Yukawa coupling $y_\chi$ can also
be expressed as the function of $\lambda$, $m_s$ and DM mass such that 
the DM mass $m_\chi$ appears as the only new additional parameter. 
As already shown in ref.~\cite{Arcadi:2016dbl}, that the presence of DM does not influence 
the prediction of 4-photon production cross-section since, when kinematically allowed, 
invisible decay processes like $s \rightarrow 
\chi \chi$ or even $a \rightarrow \chi \chi$ would feature a suppressed branching fraction.

In general a contribution from DM Yukawa $y_\chi$ should also be included 
in the RGE system, eq.~(\ref{Eq:betalambda}). 
However, being singlet under $SU(3)_C$, the related coefficients would be smaller than 
$y_F$ by a factor of three. Moreover, we always assume, throughout this work, $m_\chi \leq m_F$ 
so that $y_\chi$ is always reduced with respect to $y_F$. For this reason the impact of $y_\chi$ on the RGE 
behavior of the theory can assumed to be marginal and thus, not considered in
our analyses. 

\vspace*{0.2cm}
\noindent
{\bf{DM phenomenology:}}
For a dark matter candidate $\chi$ heavier than the pseudoscalar $a$, the correct relic density $\Omega
h^2 \simeq 0.12$~\cite{Hinshaw:2012aka,Ade:2013zuv} can be achieved through a classical freeze out
with a thermally
averaged pair annihilation cross-section  $\langle \sigma v \rangle \sim 3 \times 
10^{-26}\,{\mbox{cm}}^3 {\mbox{s}}^{-1}$. Regarding the annihilation cross-section, we can achieve good analytical 
approximations in the two limiting cases: $m_a < m_\chi < m_s/2$ and $m_\chi > m_s/2$. Notice that we will always 
assume $m_\chi < m_F$ so that the annihilation process $\chi \chi \rightarrow F F$ remains kinematically forbidden.

In the light DM regime ($m_\chi \lesssim m_s/2$)
the main annihilation channels are $\chi \chi \rightarrow gg$ and $\chi \chi \rightarrow 
\gamma \gamma$, mostly determined by the s-channel exchange of the pseudoscalar state $a$, as well as $\chi \chi 
\rightarrow aa$. The corresponding cross-sections can be approximated, using a velocity expansion, as:
\beq
\langle \sigma v \rangle_{\rm gg} =\frac{16 \lambda^2 |\widetilde{C}_{\rm GG}|^2 m_\chi^2}{\pi m_s^4} \approx 1.1 
\times 10^{-32}\,{\mbox{cm}}^3 {\mbox{s}}^{-1} {\left(\frac{\lambda}{0.01}\right)}^2 {\left(\frac{m_\chi}{100\,\mbox{GeV}}
\right)}^2 {\left(\frac{750\,\mbox{GeV}}{m_s}\right)}^4 N^2_f,
\eeq   
\beq
\langle \sigma v \rangle_{\rm \gamma \gamma} =\frac{2 \lambda^2 c_W^4 |\widetilde{C}_{\rm BB}|^2 m_\chi^2}{\pi m_s^4} \approx 2.1 
\times 10^{-34}\,{\mbox{cm}}^3 {\mbox{s}}^{-1} Q_F^4 {\left(\frac{\lambda}{0.01}\right)}^2 {\left(\frac{m_\chi}{100\,
\mbox{GeV}}\right)}^2 {\left(\frac{750\,\mbox{GeV}}{m_s}\right)}^4 N^2_f,
\eeq
\beq
\langle \sigma v \rangle_{aa}=\frac{9 m_\chi^2 \lambda^2}{384 \pi m_s^4}v^2 \approx 2.7 \times 10^{-32}
\,{\mbox{cm}}^3 {\mbox{s}}^{-1} {\left(\frac{\lambda}{0.01}\right)}^2 {\left(\frac{m_\chi}{100\,\mbox{GeV}}\right)}^2 
{\left(\frac{750\,\mbox{GeV}}{m_s}\right)}^4.
\eeq

It is evident that the annihilation cross-sections into 
the SM gauge boson pairs are s-wave dominated, i.e., do not depend on the 
velocity $v$ while the annihilation into a pseudoscalar pair is p-wave suppressed. In the 
scenario when the DM annihilation is dominated by the SM final states, the s-wave nature of the cross-section would imply 
that DM annihilations are efficient also at the present times and would be capable of generating an indirect detection 
signal, detectable with the ongoing experiments. We notice, in particular, that for $Q_F \gtrsim 2$, 
the annihilation cross-section into $\gamma \gamma$ tends to be very close to the other ones. This possibility is in 
strong tension with the limits set by FERMI~\cite{Ackermann:2015lka} gamma-ray line searches, being as strong as 
$10^{-(29 \div 30)}$ for $m_\chi \lsim 100$ GeV.

When the DM mass is well above $m_s/2$, $\langle \sigma v \rangle$ is instead dominated by the 
process $\chi \chi \rightarrow sa$ whose cross-section can be estimated as:
\beq
\langle \sigma v \rangle_{sa}=\frac{\lambda^2}{8 \pi} \frac{m_\chi^2}{m_s^4} \approx 1.2\times 10^{-29}\,{\mbox{cm}}^3 
{\mbox{s}}^{-1} {\left(\frac{\lambda}{0.01}\right)}^2 {\left(\frac{m_\chi}{500\,\mbox{GeV}}\right)}^2 {\left(\frac{750\,
\mbox{GeV}}{m_s}\right)}^4.
\eeq

This annihilation channel features a s-wave cross-section. However, the indirect signal is more peculiar since 
it would be represented by (wide) $\gamma$-ray boxes which can be investigated, in the near future, by the CTA 
experiment~\cite{Ibarra:2013eda,Ibarra:2015tya,Arcadi:2016dbl}.

Regarding the other dark matter search strategies, e.g., LHC signals like monojets~\cite{Khachatryan:2014rra,Aad:2015zva} 
are suppressed by the already mentioned low invisible branching fraction of the scalar resonance. Direct detection 
signals are similarly below the current experimental sensitivity~\cite{Akerib:2015rjg} but 
can be within the reach of upcoming future 
detectors~\cite{Aprile:2015uzo,LZ}.

As evident from the aforesaid expressions that the DM annihilation cross-section strongly depends on the parameters 
$\lambda$ and $m_s$, relevant for determining LHC observables like, 
$\sigma_{4\gamma}+\sigma_{2\gamma}$ and $\Gamma_s$, 
as well as the theoretical consistency of a model framework. As a consequence,
the requirement of the correct relic density might influence the predictions 
of the diphoton cross-section. Similarly, requiring a theoretically consistent framework up to a given energy scale 
$\Lambda_{\rm NP}$ might set limit on the viable parameter space for the DM.

By combining the aforementioned expressions of the pair annihilation 
cross-sections with eq.~(\ref{Eq:lmpi}) and eq.~(\ref{Eq:gmpi}), one 
can obtain the following approximate predictions for the 4-photon production cross-section:
\begin{align}
\label{eq:relic_vs_4gamma}
& \sigma_{4\gamma} \approx 3.4\,\mbox{pb} \,N^2_f \left(\frac{I_{\rm GG}(m_s/\sqrt{s})}{2000}\right)
{\left(\frac{\langle \sigma v \rangle_{aa}}{3\times 10^{-26}\,{\mbox{cm}}^3 {\mbox{s}}^{-1}}\right)}^{1/2}
\left(\frac{100\,\mbox{GeV}}{m_\chi}\right){\left(\frac{m_s}{1\,\mbox{TeV}}\right)}^2, \nonumber\\
& \hspace*{9.8cm} \left[\rm{for}~m_a < 3 m_{\pi^0}, m_\chi < m_s/2 \right], \nonumber\\
& \sigma_{4\gamma} \approx 1.3 \,\mbox{fb } \,N^2_fQ_F^8 \left(\frac{I_{\rm GG}(m_s/\sqrt{s})}{2000}\right)
{\left(\frac{\langle \sigma v \rangle_{aa}}{3\times 10^{-26}\,{\mbox{cm}}^3 {\mbox{s}}^{-1}}\right)}^{1/2}\left(\frac{100\,
\mbox{GeV}}{m_\chi}\right){\left(\frac{m_s}{1\,\mbox{TeV}}\right)}^2,\nonumber\\
& \hspace*{9.8cm} \left[\rm{for}~m_a > 3 m_{\pi^0}, m_\chi < m_s/2 \right],\nonumber\\
& \sigma_{4\gamma} \approx 163\,\mbox{fb} \,N^2_f \left(\frac{I_{\rm GG}(m_s/\sqrt{s})}{2000}\right){\left(\frac{\langle 
\sigma v \rangle_{sa}}{3\times 10^{-26}\,{\mbox{cm}}^3 {\mbox{s}}^{-1}}\right)}^{1/2}\left(\frac{500\,\mbox{GeV}}{m_\chi}\right)
{\left(\frac{m_s}{1\,\mbox{TeV}}\right)}^2, \nonumber\\
& \hspace*{9.8cm} \left[\rm{for}~m_a < 3 m_{\pi^0}, m_\chi > m_s/2 \right], \nonumber\\
& \sigma_{4\gamma} \approx 0.06\,\mbox{fb } \,N^2_fQ_F^8 \left(\frac{I_{\rm GG}(m_s/\sqrt{s})}{2000}\right){\left(\frac{\langle 
\sigma v \rangle_{sa}}{3\times 10^{-26}\,{\mbox{cm}}^3 {\mbox{s}}^{-1}}\right)}^{1/2}\left(\frac{500\,\mbox{GeV}}{m_\chi}\right)
{\left(\frac{m_s}{1\,\mbox{TeV}}\right)}^2, \nonumber\\
& \hspace*{9.8cm} \left[\rm{for}~m_a > 3 m_{\pi^0}, m_\chi > m_s/2 \right]. \nonumber\\
\end{align} 

We can see that in the $m_a < 3 m_{\pi^0}$ scenario, the cross-section favored by the correct relic density largely exceeds, 
almost by two orders of magnitude, the current experimental limits for $m_s$ of the order of a TeV. On the contrary, 
for $m_a > 3 m_{\pi^0}$ scenario it is possible to obtain the correct DM relic density 
compatible with an observable collimated 
photon signal for $Q_F\lesssim 2$. However, as already mentioned, for $Q_F=2$, in the low DM region the $\gamma \gamma$ 
final state channel has a too high branching fraction and thus, results in tension with limits from the indirect detection. 
We further remark that, given the $Q_F^8$ dependence of the second and fourth expression of eq.~(\ref{eq:relic_vs_4gamma}), 
the combination of the correct DM relic density and the LHC 
diphoton production cross-section\footnote{For reference, we are using a value of $\sigma_{4\gamma}$ 
of the order of a few fb.} limits imposes very stringent constraint for $Q_F \lesssim 2$.

\begin{figure}[t!]
\begin{center}
\includegraphics[width=7.5 cm]{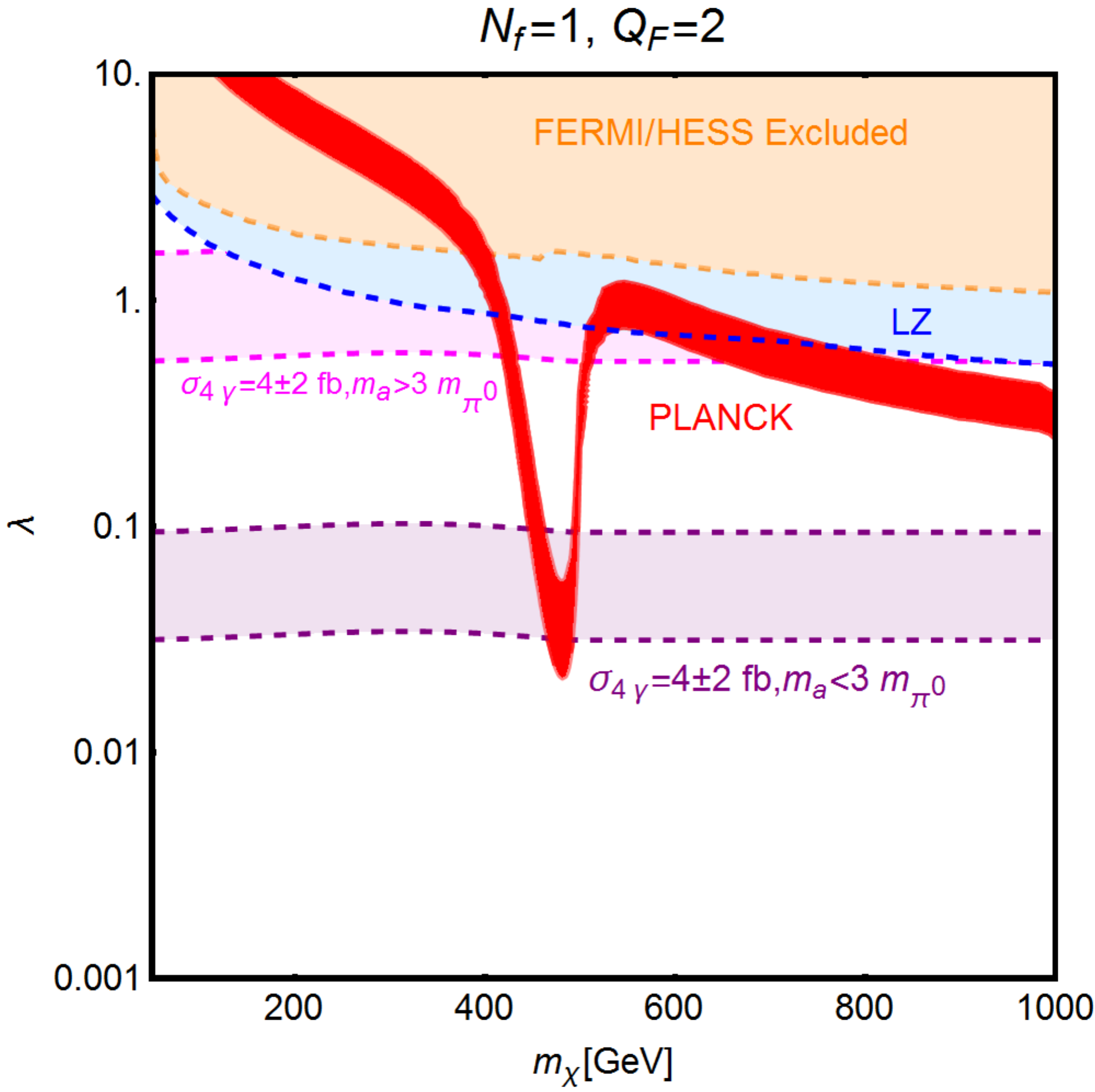}
\end{center}
\caption{\footnotesize{Isocontour (red colored) of the correct DM relic density, for $m_s=1\,\mbox{TeV},
\,N_f=1,\,Q_F=2$, in the 
$(\lambda,m_\chi)$ plane. The orange colored region is excluded by the indirect detection experiments
while the dashed blue line represents the projected LZ sensitivity \cite{LZ}. 
The magenta and purple colored bands represent the range $4 \pm 2\,\mbox{fb}$ for the collimated diphoton 
production cross-section in the $m_a < 3 m_{\pi^0}$ 
and $m_a > 3 m_{\pi^0}$ scenarios, respectively.}}
\label{fig:plot_DM}
\end{figure}

We have verified all the analytical estimates 
given in this section by numerically computing the thermally averaged DM pair annihilation 
cross-section in order to account also the pole region $m_\chi \simeq m_s/2$ properly. 
In figure~(\ref{fig:plot_DM}) we show an isocontour of 
the correct DM relic density in the bi-dimensional plane $(\lambda,m_\chi)$ 
for $N_f=1,\,Q_F=2$ and $m_s=1$ TeV. This contour is confronted with the 
prediction of a collimated diphoton cross-section, within the range of
$4 \pm 2\,\mbox{fb}$, for the two cases: $m_a > 3 m_{\pi^0}$ 
and $m_a < 3 m_{\pi^0}$. 
For the latter, as a consequence of the very low values of $\lambda$ imposed by the diphoton 
production cross-section, the only region compatible with the collider limits is the 
pole region $m_\chi \sim m_s/2$. On the contrary, in the $m_a > 3 m_{\pi^0}$ scenario, 
the suppression factor from $\mrm{Br}(a \rightarrow \gamma \gamma)^2$ should be 
compensated by a higher value of $\lambda$ which allows a fit of the DM relic density, through the annihilation 
into $sa$ final state, for DM masses of the order of 600 GeV. 
The light DM region, being excluded by limits from the Indirect Detection experiments, remains inaccessible.

It can be easily noticed that we have a more constrained scenario with respect to the one depicted
in ref.~\cite{Arcadi:2016dbl} with the effective field theory approach.
This is because, in the dynamical construction the $C_{\rm GG}$ and $C_{\rm BB}$ 
parameters as well as their relative hierarchy are not free, but substantially guided by the quantum number
assignments of the new fermions. In particular their values are always above $O(10^{-2})$, greater than the ones 
typically quoted in ref.~\cite{Arcadi:2016dbl}.

\section{Summary and discussion}
\label{Sec:summary}

In the previous sections we have described, through some examples, the relation between the collider 
limits/predictions and the theoretical consistency of the studied dynamical framework 
and their collective impact on the DM phenomenology. 
In this summary we illustrate the results of a systematic numerical study based on 
a scan involving the relevant free parameters of the model. 
The chosen parameters are the four masses $m_s,\,m_a,\,m_\chi,\,m_F$,
the electric charge $Q_F$ and the quartic coupling $\lambda$. 
The latter parameter, through eq.~(\ref{Eq:lmpi}), eq.~(\ref{Eq:gmpi}) and eq.~(\ref{2gammma1}), 
can be traded with a physical observable, i.e., the total diphoton
production cross-section $\sigma_{2\gamma}+\sigma_{4 \gamma}$. 
This collective production cross-section is varied
in the span of $10\,\mbox{fb}$, i.e., of the order of the
present experimental bounds, to $10^{-2}\,\mbox{fb}$, 
which would correspond to the observation of 
$O(1)$ events at the maximal expected luminosity for LHC run-II, $O(100\,{\mbox{fb}}^{-1})$. The four parameters 
$m_s,m_a,m_\chi,\sigma_{2 \gamma}+\sigma_{4 \gamma}$, together with $Q_F$, determine the other experimental observables, i.e., 
the total width of the scalar resonance and the interaction rates of the DM and finally, including $m_F$,  
the energy scale $\Lambda_{\rm NP}$ at which the theory should be UV completed.

Choosing one family of the new BSM fermion, i.e., $N_f=1$,
we have varied the three masses $m_s$, $m_a$, $m_\chi$ and $\sigma_{2 \gamma}+\sigma_{4 \gamma}$ in the following ranges:
\bea
&& m_s \in \left[200,2000\right]\,\mbox{GeV}, \nonumber\\
&& m_a \in \left[0.2,2\right]\,\mbox{GeV},\nonumber\\
&& m_\chi \in \left[100,1000\right]\,\mbox{GeV},\nonumber\\
&& \sigma_{2 \gamma}+\sigma_{4 \gamma} \in \left[10^{-2},10\right]\,\mbox{fb}.
\eea 

The scan has been repeated for five discrete values of $m_F$, namely $1,2,3$ and 
$5\,\mbox{TeV}$ and for $Q_F=1$ and $2$. For each set of the 
considered parameters we have computed the LHC and DM observables as well as the value of 
$\Lambda_{\rm NP}$. Further, for each model configuration we have imposed 
the constraints of correct DM relic density\footnote{In our numerical
study we have estimated the relic density by imposing a bound on
the pair annihilation cross-section in the span of $2\times 10^{-26}$ to 
$5 \times 10^{-26}~\rm{cm}^3 ~\rm{s}^{-1}$.}, observational DM limits, a pseudoscalar 
decay length $l \leq 1\,\mbox{m}$ and finally, $\Gamma_s/m_s \leq 10\%$.\
Furthermore, we consider $m_F \geq m_s$ and $y_F < \sqrt{4\pi}$ 
in all our analyses.

A first outcome of this scan has already been depicted in figure~\ref{Fig:lambdanp}, which shows that for $m_F=3$
and $5$ TeV it is extremely difficult, or even impossible, to achieve observable diphoton production
cross-section in a theoretically consistent way. Thus, we will focus our discussion 
on the results of the cases $m_F=1$ and $2\,\mbox{TeV}$.

\begin{figure}[t!]
\begin{center}
\includegraphics[width=7.9 cm]{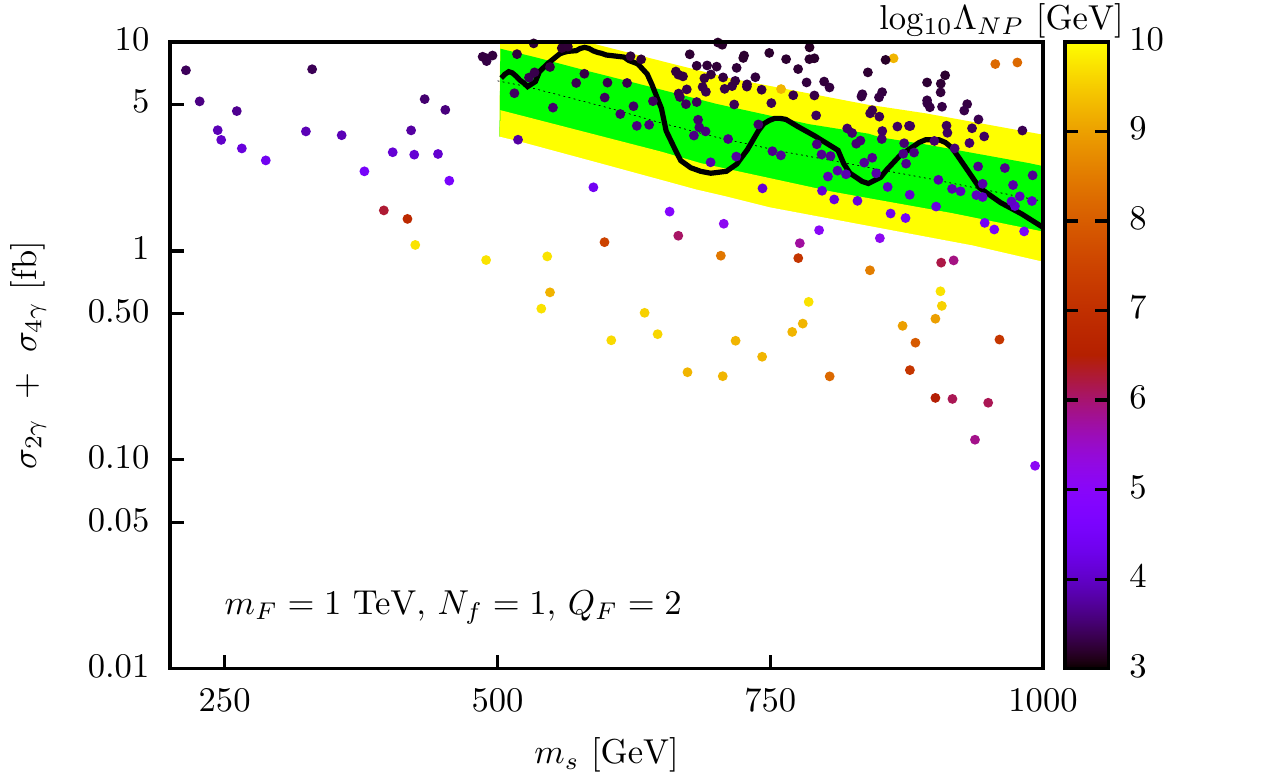}
\hspace*{-0.70cm}
\includegraphics[width= 7.9 cm]{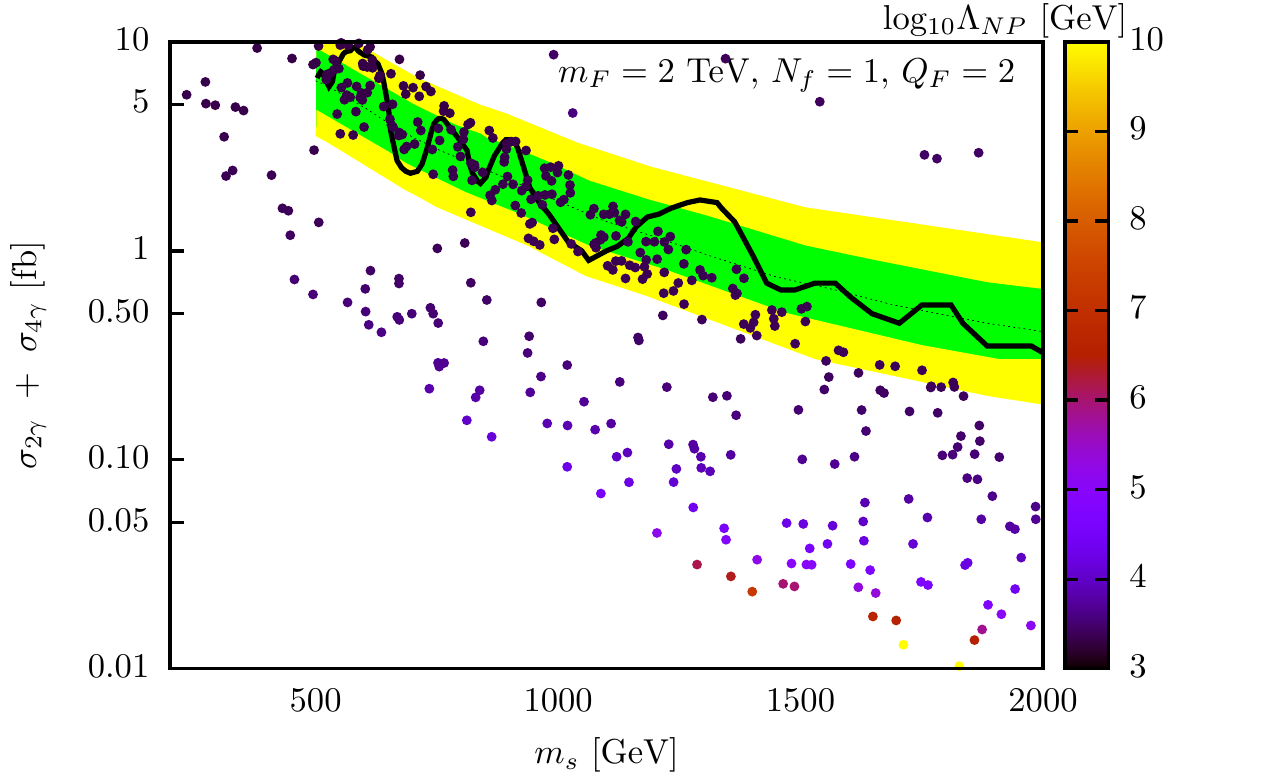}\\
\includegraphics[width=7.9 cm]{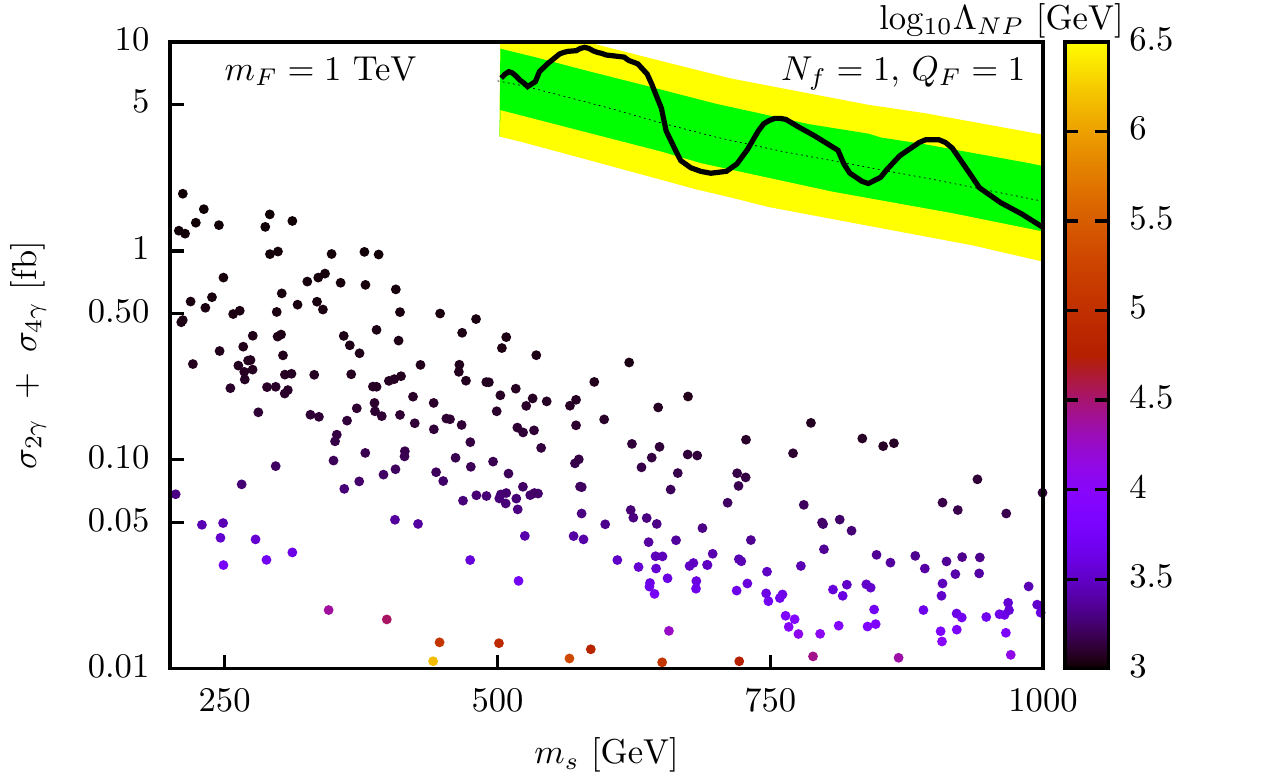}
\hspace*{-0.70cm}
\includegraphics[width= 7.9 cm]{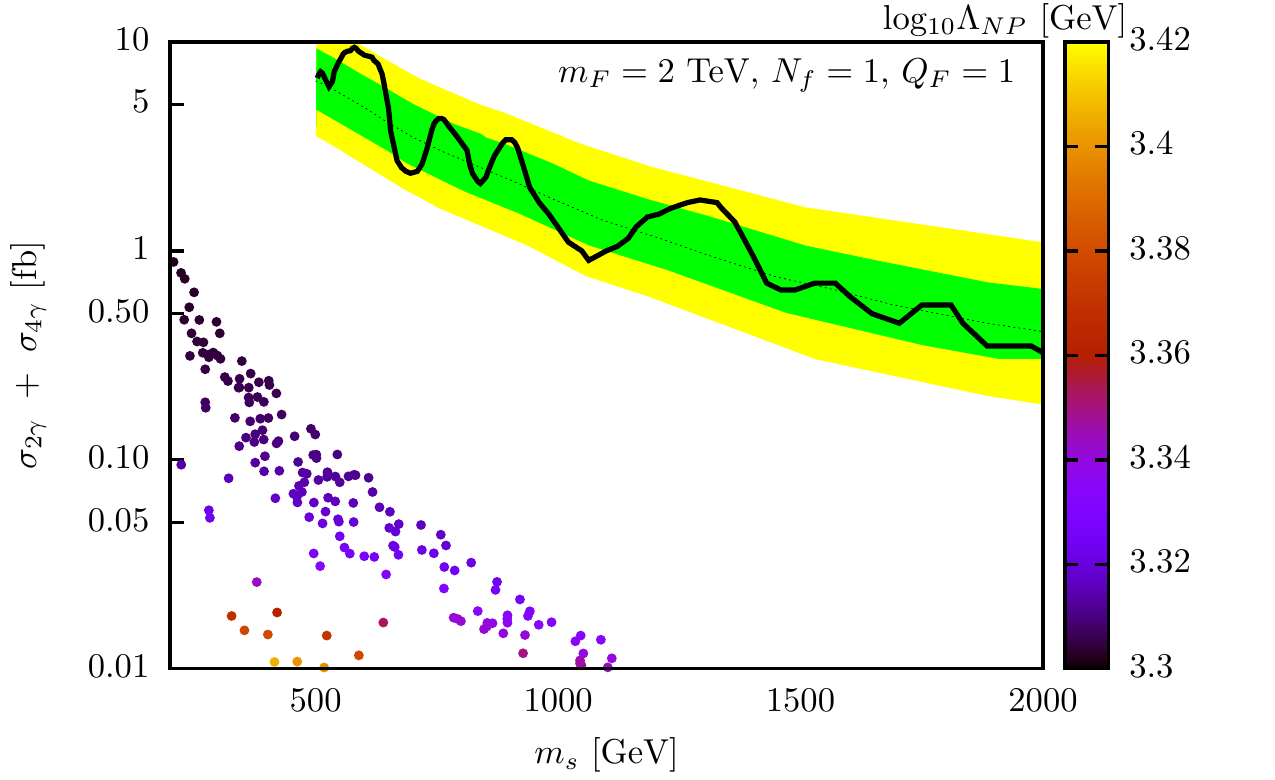}
\end{center}
\caption{\footnotesize{Distribution of the model configurations in the 
$(\sigma_{2\gamma}+\sigma_{4\gamma},m_s)$ plane. The corresponding
$m_F$ and $Q_F$ values are given in the plots. The color
coding represents the different values of $\Lambda_{\rm NP}$.
The experimental data (represented with the familiar
yellow-green plot) used in these plots corresponds
to 16.2 fb$^{-1}$ of data at $\sqrt{\bf{s}}$=13 TeV 
+ 19.7 fb$^{-1}$ of data at $\sqrt{\bf{s}}$=8 TeV,
with $\Gamma_s/m_s=5.6\%$, from the latest CMS result \cite{CMSICHEP16}.}}
\label{fig:temperature1}
\end{figure}

We have plotted in figure \ref{fig:temperature1} 
the prediction of a diphoton signal at the LHC, i.e., $\sigma_{2\gamma}+\sigma_{4\gamma}$,
as a function of the mass of the scalar resonance, where each of the model 
configurations are tested against the list of constraints mentioned above. The associated
color coding represents the value of $\Lambda_{\rm NP}$ for a given model configuration.
The predictions, estimated with this model, are compared\footnote{A judicious comparison
should also include issues like the detection efficiency, acceptance against
the applied cuts, scaling for higher order effects, etc. Such intricate details
are beyond the scopes of the current analysis.}
with the limits (plotted with the familiar yellow-green style) recently reported by the CMS group~\cite{CMS1}.
From figure~\ref{fig:temperature1} it is apparent, that for $Q_F=2$ (top row) most of the points lie within the 
current experimental sensitivity or slightly below, especially when $\Lambda_{\rm NP}$ is large.
The region with large $\Lambda_{\rm NP}$, as expected from the UV behavior of the theory (see section \ref{Sec:UV}),
shifts towards higher $m_s$ values for increasing $m_F$ and 
produces an anticipated reduction in $\sigma_{2\gamma}+\sigma_{4\gamma}$.

Further, the configurations lying within the 1-2$\sigma$ bands 
of the current experimental observation typically appear with the lower 
values of $\Lambda_{\rm NP}$. This is because, for a given value of $m_s$, a higher production 
cross-section implies higher values of $\lambda$ (see subsection \ref{subsec:4gamma}) 
and hence, a higher starting values of $y_F$ since $y_F\propto \sqrt{\lambda}$.
This, in turn would produce a more negative $\beta_\lambda/\lambda$,
forcing $\lambda$ to turn negative at a lower $\Lambda_{\rm NP}$.
Relatively well UV behaved configurations lie below the current 
experimental limits and are expected to be probed in the near future.

For $Q_F=1$, on the contrary, all the points lie well below the current
experimental sensitivity. This is in agreement with eq.~(\ref{eq:relic_vs_4gamma}) which 
suggests that the DM relic density favors very small, even below $10^{-2}\,\mbox{fb}$ 
LHC cross-sections. We also notice that in the $Q_F=1$ case (bottom row of 
figure~\ref{fig:temperature1}), a much stronger requirement exists on the value 
of $m_F$ from the UV behavior of the theory. We observe that basically all the points, 
for $m_F=2\,\mbox{TeV}$, feature $\Lambda_{\rm NP} 
\simeq m_F$ (bottom row, right) and in the case of $m_F=1\,\mbox{TeV}$, one gets 
$\Lambda_{\rm NP}$ not exceeding $10^6\,\mbox{GeV}$ (bottom row, left). 
This arises due to the $Q_F^8$ dependence of the 4-photon production cross-section 
(we remind that the regime $m_a < 3 m_{\pi^0}$, for which the cross-section does not 
depend on $Q_F$, is largely incompatible with the correct relic density) which 
implies larger values of $\lambda$ for obtaining the same value of the production cross-section (see 
also figure~\ref{fig:magL}). The large $\lambda$ values would lead to a more problematic UV behavior of the model. 


\begin{figure}[t!]
\begin{center}
\includegraphics[width=7.9 cm]{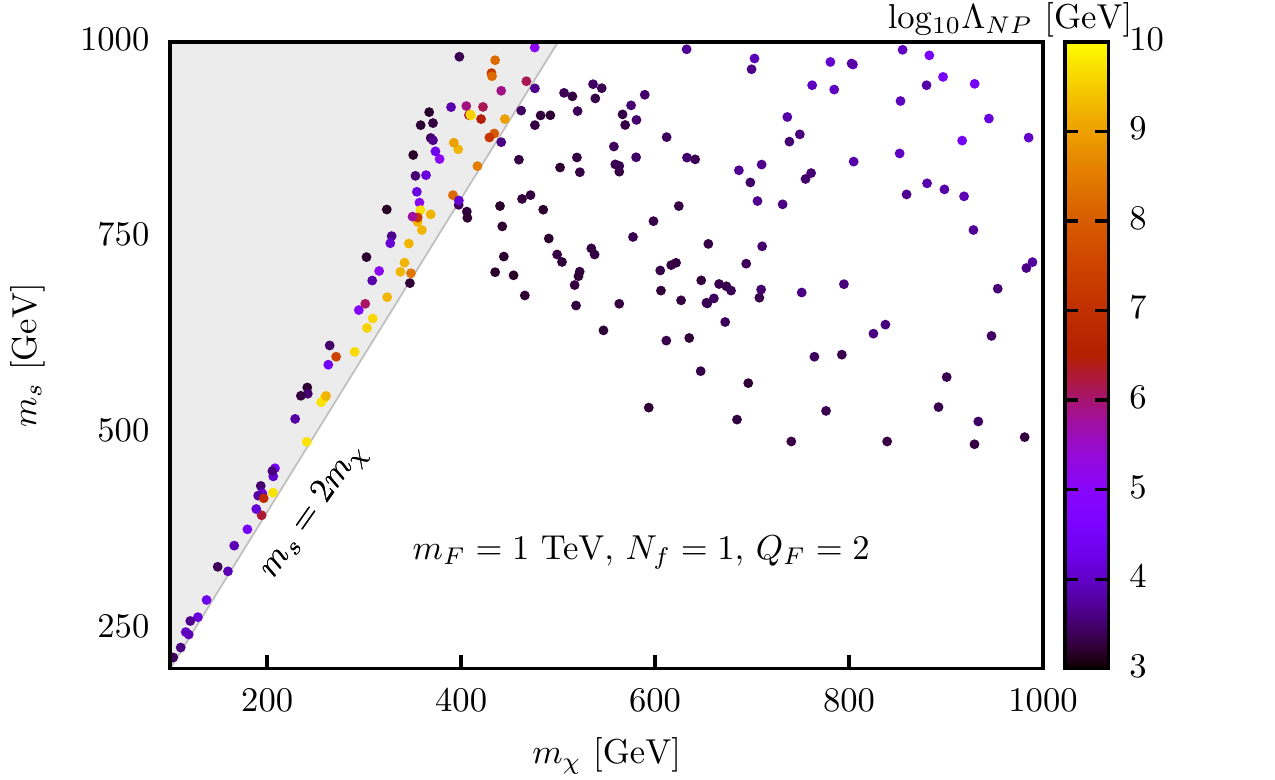}
\hspace*{-0.75cm}
\includegraphics[width= 7.9 cm]{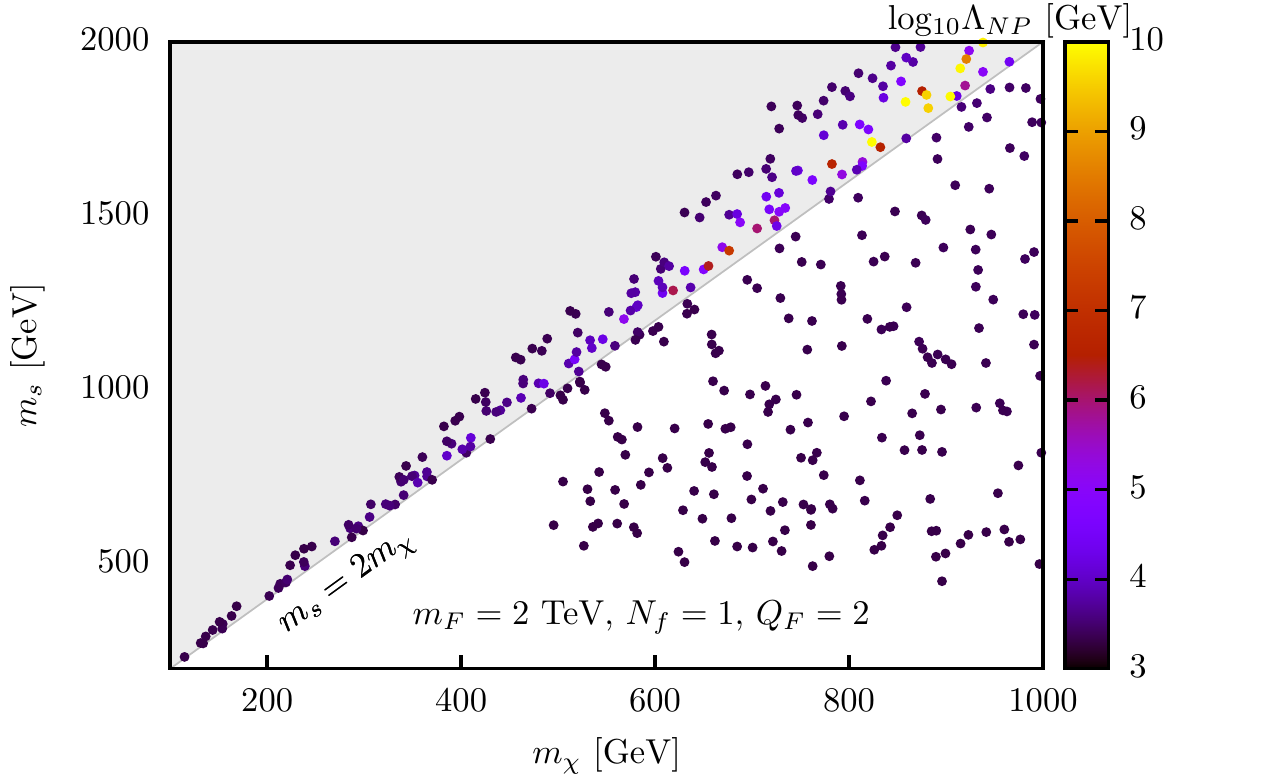}
\end{center}
\caption{\footnotesize{Distribution of the model configurations in the $(m_\chi,m_s)$ plane
for the different $m_F$, $Q_F$ values, mentioned in the plots. The gray
colored region represents the $m_\chi \leq m_s/2$ scenario.
The color coding represents the different values of $\Lambda_{\rm NP}$.}}
\label{fig:temperature2}
\end{figure}

Finally, just for the $Q_F=2$ scenario, we have shown in figure \ref{fig:temperature2} 
the distribution of points in the $(m_\chi,m_s)$ plane to 
highlight the regions of the parameter space favored by the correct DM relic density. 
Once again, as suggested by eq.~(\ref{eq:relic_vs_4gamma}), we notice that the region corresponding 
to $m_\chi < m_s/2$ (gray colored) is substantially empty. 
Regarding the value of $\Lambda_{\rm NP}$, we observe
that the maximal values lie in the resonance region 
$m_\chi \simeq m_s/2$. This happens since the enhancement of the DM annihilation cross-section 
in this regime allows to fit the correct relic density for lower values of $\lambda$ 
which guarantee a better UV behavior of the theory.

\section{Conclusion}
\label{Sec:conclu}

In this article we have presented a model where the SM is extended 
by a SM gauge singlet complex scalar field $\Phi$, a set of new BSM fermions 
charged also under the SM gauge group and a fermionic DM candidate, singlet under the SM 
gauge charges but charged under the new global $U(1)$ symmetry associated with the $\Phi$ field. 
All the masses in this theory, except for the pseudo Goldstone $a$,
are dynamically generated from spontaneous breaking of the associated $U(1)$ symmetry so that the theory has 
only one fundamental scale, represented by the scale of spontaneous symmetry breaking.

The typical LHC signature of this model is the production of pairs of collimated photons produced from the 
decay of a pair of pseudo Goldstone bosons $a$ (the imaginary component of $\Phi$), originated 
from a scalar resonance $s$ (the real component of $\Phi$) while other possible processes, like dijets, 
$Z\gamma$ and $ZZ$ productions, have rather suppressed rates. The couplings responsible for the 
collective diphoton, i.e., $2\gamma+4\gamma$, 
production are induced by the new BSM fermions at the one-loop level. However, given the dynamical structure 
of the model, the couplings are practically independent of the BSM fermion masses, so that the cross-section depends mainly
on $m_s$, the quartic coupling $\lambda$, the number of BSM fermion families $N_f$
and on their electric charges $Q_F$. As a consequence there exists 
direct correlation between the collective diphoton production cross-section, 
i.e., $\sigma_{2\gamma}+\sigma_{4\gamma}$, and the total decay width $\Gamma_s$ of the resonance.

The presence of colored and electrically charged new BSM fermions has a strong impact on the 
UV behavior of the theory. Indeed the quartic coupling $\lambda$ can be driven rather efficiently towards negative 
values at a energy scale $\Lambda_{\rm NP}$ in the vicinity of $m_F$ such that a further completion of the 
theory should be invoked at the energy scale $\Lambda_{\rm NP}$. Again, the dynamical structure of the 
model gives a tight relation between the value of $\Lambda_{\rm NP}$ and the experimentally 
measurable diphoton production cross-section.

We have shown that for masses of the new BSM fermions not exceeding 2 TeV, having electric
charge $Q_F=1$ or $2$, the considered framework is theoretically consistent 
up to some energy scale $\Lambda_{\rm NP} \sim 2.5~{\rm TeV}~(10^{10}~{\rm GeV})$ for $Q_F=1~(2)$,
starting from $ \gtrsim m_F$. Thus, at least part of the $\Lambda_{\rm NP}$ remains within the reach
of ongoing and upcoming collider experiments and,
at the same time can produce an observable signal. It is possible,
in addition, to achieve a DM candidate with the correct relic density without conflicting with 
the existing search results. 
The DM sector for this model is capable of producing signals in the indirect 
detection experiments which are expected to be probed in the near future.

\acknowledgments
Y.M. wants to thank especially J.B de Vivie whose help was fundamental throughout our work
The authors thanks Emilian Dudas and Ulrich Ellwanger for fruitful discussions. 
This project has received funding from the European Union's Horizon 2020
research and innovation programme under the Marie Sk\l{}odowska-Curie grant
agreements No. {\bf 690575} and  No. {\bf 674896}.
This work is also supported by the Spanish MICINN's Consolider-Ingenio 2010 
Programme under grant Multi-Dark {\bf CSD2009-00064}, 
the contract {\bf FPA2010-17747}, the France-US PICS no. 06482 and the LIA-TCAP of CNRS. 
Y.~M. and G. A. acknowledges partial support the ERC advanced grants Higgs@LHC and MassTeV. 
G. A. thanks the CERN theory division for the hospitality during
part of the completion of this project.
P.G. acknowledges the support from P2IO Excellence Laboratory (LABEX). 
This research was also supported in part by the Research
Executive Agency (REA) of the European Union under
the Grant Agreement {\bf PITN-GA2012-316704} (``HiggsTools'').

\appendix
\section{Loop induced coefficients}

For completeness, in this appendix we present the detailed  
expressions for the loop induced coefficients 
$C_{\rm BB}$, $C_{\rm GG}$, $\widetilde{C}_{\rm BB}$ and $\widetilde{C}_{\rm GG}$.

Given the effective Lagrangian:
\beq
-{\cal L_\mrm{eff}} \supset k_{GG} s G^\alpha_{\mu \nu} G^{\mu \nu}_\alpha
+k_{BB} s B_{\mu \nu} B^{\mu \nu}+ \tilde{k}_{GG} a
G^a_{\mu \nu} \widetilde G^{\mu \nu}_a
+\tilde{k}_{BB} a  B_{\mu \nu} \widetilde B^{\mu \nu},
\eeq
its dimension-full coefficients, when they are originated from the one-loop triangle diagrams
involving new fermionic states of masses $m_{F_\alpha}$, can be written as~\cite{Bae:2016xni}:

\beq
k_{GG}=\frac{\alpha_s}{8\pi}\left( \sum_\alpha d_\alpha^{(2)} C_\alpha^{(3)}\frac{1}{\sqrt{2}} 
\frac{y_{F_\alpha}}{m_{F_\alpha}} f_{1/2}(\tau_\alpha) \right),
\eeq
\beq
k_{BB}=\frac{\alpha_{\text{em}}}{8\pi}\frac{1}{c_W^2}\left( \sum_\alpha (d_{\alpha}^{(3)} C_{\alpha}^{(2)}
+d_{\alpha}^{(3)} d_{\alpha}^{(2)}Q^2_{F_\alpha})\frac{1}{\sqrt{2}} 
\frac{y_{F_\alpha}}{m_{F_\alpha}}f_{1/2}(\tau_\alpha) \right),
\eeq
for the scalar and for the pseudoscalar:
\beq
\tilde{k}_{GG}=\frac{\alpha_s}{8\pi}\left( \sum_\alpha d_\alpha^{(2)} C_\alpha^{(3)} 
\frac{1}{\sqrt{2}}\frac{y_{F_\alpha}}{m_{F_\alpha}} \tilde{f}_{1/2}(\tau_\alpha) \right),
\eeq
\beq
\tilde{k}_{BB}=\frac{\alpha_{\text{em}}}{8\pi}\frac{1}{c_W^2}\left( \sum_\alpha (d_{\alpha}^{(3)} 
C_{\alpha}^{(2)}+d_{\alpha}^{(3)} d_{\alpha}^{(2)}Q^2_{F_\alpha})\frac{1}{\sqrt{2}} \frac{y_{F_\alpha}}{m_{F_\alpha}}
\tilde{f}_{1/2}(\tau_\alpha) \right),
\eeq
where loop induced $f_{1/2}(\tau_\alpha)$ and ${\tilde{f}}_{1/2}(\tau_\alpha)$ functions are defined as:
\beq
f_{1/2}(\tau)=\frac{2 (\tau+(\tau-1) f(\tau))}{\tau^2},\,\,\,\,\tilde{f}_{1/2}(\tau)=2 \frac{f(\tau)}{\tau},
\eeq
with:
\beq
f(\tau)=
\left \{
\begin{array}{cc}
arc\sin^2(\sqrt{\tau}) & \tau \leq 1 \\
-\frac{1}{4}{\left(\log\left[\frac{1+\sqrt{1-1/\tau}}{1-\sqrt{1-1/\tau}}\right]\right)}^2 & \tau >1,
\end{array}
\right.
\eeq
and $\tau_\alpha$ defined as :
\beq
\tau_\alpha=\frac{m_s^2}{4 m^2_{F_\alpha}}~\mrm{(for~ scalar})\,\,\,\rm{and}\,\,\,
\tau_\alpha=\frac{m_a^2}{4 m^2_{F_\alpha}}~\mrm{(for~ pseudoscalar)}.
\eeq

For $SU(N)$ symmetry group we have  $(d_\alpha^{(N)},C_\alpha^{(N)})=(1,0),\,(N,1/2)$ and $(N^2-1,N)$ for 
a singlet, fundamental and adjoint representation, respectively. Since we have considered the addition of mass 
degenerate fermions with same quantum number assignments, apart from the sign of $Q_F$, 
we can straightforwardly replace 
$\sum_\alpha \rightarrow 2 N_f$. Furthermore, in our dynamical setup $y_F$ and $m_F$ are not independent parameters 
but are related as $y_F=2 \sqrt{\lambda} ({m_F}/{m_s})$. Thus,
apart from the loop functions $f_{1/2}(\tau_\alpha)$ and ${\tilde{f}}_{1/2}(\tau_\alpha)$,
it is possible to eliminate the $m_F$ dependence in $k_{GG},\,k_{BB}$ etc.
The coefficients of eq.~(\ref{eq:coeff}) are obtained by setting $k_{VV} (\tilde{k}_{VV})=
\frac{\sqrt{\lambda}}{m_s} C_{VV}(\widetilde{C}_{VV})$ where $V=B,\,G$.

\bibliography{AGMP2ref1v6}

\end{document}